\documentclass[fleqn,usenatbib]{mnras}

\usepackage{mathptmx}

\usepackage[T1]{fontenc}
\usepackage{ae,aecompl}

\usepackage{graphicx}	
\usepackage{amsmath}	
\usepackage{amssymb}	
\usepackage{color}
\usepackage{longtable}
\usepackage{pdflscape}

\title[YSO variability in the LMC]{The VMC survey -- XXXVI. Young stellar variability in the Large Magellanic Cloud}

\author[V. Zivkov et al.]
{Viktor Zivkov,$^{1,2}$\thanks{E-mail: v.zivkov@keele.ac.uk}
Joana M. Oliveira,$^{1}$
Monika G. Petr-Gotzens,$^{2,12}$
\newauthor
Stefano Rubele,$^{3,4}$
Maria-Rosa L. Cioni,$^{5}$
Jacco Th. van Loon,$^{1}$
\newauthor
Richard de Grijs,$^{6,7,8}$
Jim Emerson,$^{9}$
Valentin D. Ivanov,$^{2}$
Marcella Marconi,$^{10}$
\newauthor
Maria Ida Moretti,$^{10}$
Vincenzo Ripepi,$^{10}$
Florian Niederhofer$^{5}$
and Ning-Chen Sun$^{11}$
\\
$^{1}$Lennard-Jones Laboratories, School of Chemical and Physical Sciences,
   		Keele University, ST5 5BG, UK\\
$^{2}$European Southern Observatory, Karl-Schwarzschild-Str. 2,
             85748 Garching bei M\"unchen, Germany\\
$^{3}$Dipartimento di Fisica e Astronomia, Universit\`a di Padova,
         	Vicolo dell'Osservatorio 2, I-35122 Padova, Italy\\
$^{4}$Osservatorio Astronomico di Padova -- INAF,
            Vicolo dell'Osservatorio 5, I-35122 Padova, Italy\\
$^{5}$Leibniz-Institut f\"ur Astrophysik Potsdam, An der Sternwarte 16, D-14482                 Potsdam, Germany\\
$^{6}$Department of Physics and Astronomy, Macquarie University, Balaclava Road,                Sydney NSW 2109, Australia\\
$^{7}$Centre for Astronomy, Astrophysics and Astrophotonics, Macquarie University,              Balaclava Road, Sydney NSW 2109, Australia\\
$^{8}$International Space Science Institute--Beijing, 1 Nanertiao, Zhongguancun,                Hai Dian District, Beijing 100190, China\\
$^{9}$Astronomy Unit, School of Physics and Astronomy, Queen Mary University of                 London, Mile End Road, London E1 4NS, UK\\
$^{10}$INAF-Osservatorio Astronomico di Capodimonte, via Moiariello 16,                        I-80131, Naples, Italy\\
$^{11}$Department of Physics and Astronomy, University of Sheffield, Hicks Building, Hounsfield Road, Sheffield S3 7RH, UK\\
$^{12}$Universit\"ats-Sternwarte, Ludwig-Maximilians-Universit\"at M\"unchen, Scheinerstr 1, D-81679 M\"unchen, Germany
}

\date{Accepted XXX. Received YYY; in original form ZZZ}

\pubyear{2015}

\begin{document}
\label{firstpage}
\pagerange{\pageref{firstpage}--\pageref{lastpage}}
\maketitle

\begin{abstract}

Studies of young stellar objects (YSOs) in the Galaxy have found that a significant fraction exhibit photometric variability. However, no systematic investigation has been conducted on the variability of extragalactic YSOs. Here we present the first variability study of massive YSOs in a $\sim 1.5\,\mathrm{deg^2}$ region of the Large Magellanic Cloud (LMC). The aim is to investigate whether the different environmental conditions in the metal-poor LMC ($\sim 0.4-0.5\,\mathrm{Z_{\sun}}$) have an impact on the variability characteristics. Multi-epoch near-infrared (NIR) photometry was obtained from the VISTA Survey of the Magellanic Clouds (VMC) and our own monitoring campaign using the VISTA telescope. By applying a reduced $\chi^2$-analysis, stellar variability was identified. We found 3062 candidate variable stars from a population of 362\,425 stars detected. Based on several \textit{Spitzer} studies, we compiled a sample of high-reliability massive YSOs: a total of 173 massive YSOs have NIR counterparts (down to $K_{\mathrm{s}}\sim 18.5\,$mag) in the VMC catalogue, of which 39 display significant ($>3\sigma$) variability. They have been classified as eruptive, fader, dipper, short-term variable and long-period variable YSOs based mostly on the appearance of their $K_{\mathrm{s}}$ band light curves. The majority of YSOs are aperiodic, only five YSOs exhibit periodic lightcurves. The observed amplitudes are comparable or smaller than those for Galactic YSOs (only two Magellanic YSOs exhibit $\Delta K_{\mathrm{s}}>1\,$mag), not what would have been expected from the typically larger mass accretion rates observed in the Magellanic Clouds.
\end{abstract}




\section{Introduction}
\label{sec:intro}

Photometric variability was early on recognised as one of the defining characteristics of young stars prior to their arrival on the main sequence \citep{Joy1945, Herbig1952}. In one of the first near-infrared (NIR) monitoring programmes of young stars \citep{Skrutskie1996}, all pre-main sequence (PMS) stars in the sample exhibited statistically significant variability. Subsequent large sample programmes showed that 50--60\% of all young stars are variable \citep{Carpenter2001, Morales2011}, making variability an excellent tracer of stellar youth \citep[$\lesssim 10\,$Myr;][]{Briceno2005}. The large range of variability  time-scales (days to years), amplitudes (one tenth to over two magnitudes) and light curve shapes (periodic and sinusoidal, periodic non-sinusoidal, irregular) suggests a variety of physical mechanisms leading to variability.

Different variability patterns can be associated with different mechanisms such as rotational modulation of hot and cool star spots, obscuration by disc structures like warps or clumps and unsteady mass accretion \citep[e.g.][]{Cody2014, Rice2015}. Galactic studies like the Young Stellar Object (YSO) Variability \citep[YSOVAR;][]{Rebull2014} programme found clear correlations between {\it global} variability characteristics and the age of a young stellar population.  Long variability timescales (weeks or longer) and large amplitudes of $\Delta K_{\mathrm{s}}>1$\,mag dominate amongst stars in early evolutionary stages \citep[e.g. Class I protostars;][]{Contreras2014}, likely as a result of unsteady mass accretion. More evolved PMS stars show on average smaller variability amplitudes (usually $\Delta K_{\mathrm{s}}<0.5\,$mag) and shorter periods ($P\lesssim\,15\,$d), caused by obscuring structures in the inner circumstellar disc (for Class II objects), or by photospheric phenomena like cool and/or hot spots (for Class III and Class II sources). Moreover, PMS intermediate mass stars ($\sim 1-4\,\mathrm{M_{\sun}}$) also show $\delta$ Scuti-like intrinsic pulsations when crossing the classical instability strip of more evolved pulsating stars (see \citealt{Marconi1998, Zwintz2009} and references therein).

\textit{Hubble Space Telescope} (HST) observations of star forming regions in the metal-poor Large and Small Magellanic Clouds (LMC, SMC; $Z_{\mathrm{LMC}}\approx 0.4-0.5\,\mathrm{Z_{\sun}}$, \citealt{Choudhury2016}; $Z_{\mathrm{SMC}}\approx 0.15\,\mathrm{Z_{\sun}}$, \citealt{Choudhury2018}) have reported consistently higher mass accretion rates (as probed by H$\alpha$ emission) compared with Galactic samples of PMS stars of similar age and mass \citep{Marchi2011,Marchi2013,Spezzi2012}. Higher mass accretion rates for massive YSOs are also reported by \citet{Ward2016,Ward2017}. 
As the mass accretion rates scale with circumstellar disc mass \citep[e.g.][]{Manara2016}, this implies higher disc masses for Magellanic young stars, likely enabled by weaker radiation pressure as a result of the higher gas-to-dust ratio. More massive discs are prone to gravitational instabilities \citep{Evans2015} which can cause strong variability in the inward mass accretion, potentially leading to larger variability amplitudes among Magellanic YSOs compared with Galactic samples. Other studies indicate that disc lifetimes decrease with lower metallicity \citep{Ercolano2010,Yasui2010,Yasui2016}, in which case more short-period variables with smaller amplitudes would be expected.
Stars with variability caused by obscuration events from a dusty disc likely display on average smaller amplitudes than similar stars in the Galaxy because of the lower disc opacity \citep{Durisen2007}.

\citet{Vijh2009} examined the mid-infrared variability of LMC stars using data obtained by the \textit{Spitzer Space Telescope} observing programme SAGE \citep[Surveying the Agents of a Galaxy's Evolution;][]{Meixner2006}. They found $\sim 2000$ variables which were mostly evolved asymptotic giant branch (AGB) stars. Cross-correlating these variables with the list of $\sim 1000$ YSOs from \citet{Whitney2008} revealed that 29 variables are likely YSOs, resulting in a YSO variability fraction of around 3\%. However, with only two epochs available in the SAGE catalogues it is not possible to constrain amplitudes and periods. The SAGE-var study \citep{Riebel2015} added four observational epochs to the SAGE data; using the same criteria as \citet{Vijh2009}, it identified 2198 variables in the LMC, of which only 12 are YSO candidates. These small numbers either indicate a very low sensitivity to young variables or alternatively a low variability fraction amongst young stars.

The VISTA Survey of the Magellanic Clouds \citep[VMC; ][]{Cioni2011} provides up to 12 epochs of observations in the $K_{\mathrm{s}}$ band. Analysis of variable sources based on VMC data was presented in more than ten VMC papers; for planetary nebulae \citep{Miszalski2011}, quasars \citep{Cioni2013, Ivanov2016} and pulsating variable stars \citep[e.g.][]{Ripepi2012, Ragosta2019}. In particular, \citet{Cioni2013} and \citet{Moretti2016} presented new selection criteria for variable sources for quasars and Cepheids, respectively.

By combining the VMC epochs and those obtained by an open time European Southern Observatory (ESO) programme, we obtained up to 25 epochs for a $\sim 1.5\,\mathrm{deg^2}$ region in the LMC (VMC tile LMC 7\_5; see Fig.\,\ref{fig:LMC75}). In this work we probe the observed light curves for variability using a reduced $\chi^2$-analysis. A total of 3062 candidate variables are identified. Starting with samples of \textit{Spitzer}-identified massive YSOs we characterise  the NIR light curves of 173 high reliability YSOs. Of these, 39 YSOs are found to be variable and their light curves are classified based on shape and periodicity.

The paper is organised as follows. Sect.\,\ref{sec:data} provides an overview of the data used in this work, and Sect.\,\ref{sec:source_tracking} explains how the multi-epoch catalogue was constructed. In Sect.\,\ref{sec:chi2_test} we describe the identification criteria for photometric variability. The reliability of our method and its ability to recover known periods is assessed in Sect.\,\ref{sec:analysis_varstars} based on a sample of OGLE variable stars \citep{Udalski2008}.  In Sect.\,\ref{sec:YSO_sample} we describe the massive YSO sample selection and their NIR counterparts in the VMC point spread function (PSF) photometric catalogues. We present and discuss the results of the YSO variability analysis in Sect.\,\ref{sec:YSO_varresults}, and conclude with a summary in Sect.\,\ref{sec:YSO_var_conc}.


\section{Near-infrared data}
\label{sec:data}
We combined the data obtained by the VMC survey with observations from the ESO open time programme 0100.C-0248(A). This extends the time baseline to a total of $\sim 6$\,yr, and adds short cadence observations. Both programmes obtained NIR observations using the 4.1\,m Visible and Infrared Survey Telescope for Astronomy \citep[VISTA;][]{Sutherland2015}.

\subsection{VMC data}
\label{subsec:VMC_data}
The VMC is a large ESO public survey that observed an area of $\sim170\,\mathrm{deg^2}$ in the Magellanic System, with a total of 110 tiles observed. Each tile covers uniformly an area of $\sim$\,1.5\,$\mathrm{deg^2}$ as a result of combining six offset pawprint images in order to fill the gaps between the 16 VIRCAM detectors. The observing strategy of the VMC involves multi-epoch imaging of tiles in the $Y$ ($1.02\,\mathrm{\mu m}$), $J$ ($1.25\,\mathrm{\mu m}$) and $K_{\mathrm{s}}$ ($2.15\,\mathrm{\mu m}$) bands. Every tile is observed in two deep epochs in both $Y$ and $J$, and 11 deep epochs in $K_{\mathrm{s}}$; the exposure time per pixel and epoch is 800\,s in $Y$ and $J$, and 750\,s in $K_{\mathrm{s}}$. In addition, there are two shallow epochs per band with half the exposure time. This results in total integration times of 2400\,s ($Y$, $J$) and 9000\,s ($K_{\mathrm{s}}$) in the deep stacked image of all epochs. The nominal $10\sigma$ magnitude limits are $Y\approx 21.9$\,mag, $J\approx 21.4$\,mag and $K_{\mathrm{s}}\approx 20.3$\,mag (Vega system). 

Pawprint images, spanning roughly 4 years of VMC observations, were reduced and calibrated with the VISTA Data Flow System (VDFS) pipeline v1.3 \citep{Irwin2004, Gonzales2018}. Pawprint PSF photometry was obtained using IRAF Daophot tasks. Photometric errors were estimated using a combination of background-noise error and PSF model fitting error.

\begin{figure}
	\includegraphics[width=\columnwidth]{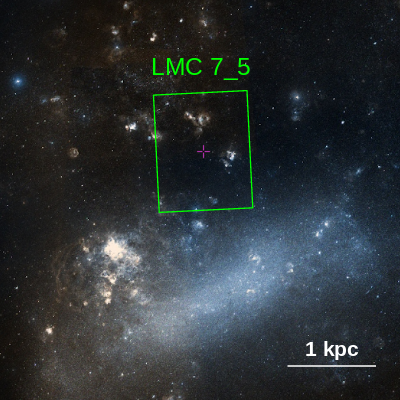}
    \caption{Digitized Sky Survey (DSS) image showing the location of tile LMC 7\_5 (highlighted in green) within the wider LMC environment. North is up and east is to the left.}
    \label{fig:LMC75}
\end{figure}

The region covered by tile LMC 7\_5 (Fig.\,\ref{fig:LMC75}) hosts large star forming complexes \citep[N44 and N51, e.g.][]{Carlson2012}, and VMC observations uncovered rich PMS populations \citep{Zivkov2018}. The tile central coordinates are $\alpha\,\mathrm{(J2000)} \approx 81\fdg493$ and $\delta\,\mathrm{(J2000)} \approx -67\fdg895$. The exposure times per pawprint are 400\,s in the $Y$ and $J$ bands, and 375\,s in the $K_{\mathrm{s}}$ band. 
Since seeing varies from epoch to epoch the completeness also changes. Sources at around $J\approx 20.5$\,mag and $K_{\mathrm{s}}\approx 18.4$\,mag are typically detected in half of the pawprints they are detected in. Appendix \ref{app:A} contains detailed listings of all epochs and pawprints (Tables \ref{tab:Kepochs_data} and \ref{tab:Jepochs_data}).

\subsection{Complementary data}
\label{subsec:complementary_data}
We complement the VMC epochs in the $J$ and $K_{\mathrm{s}}$ bands with observations of tile LMC 7\_5 obtained by the open time programme 0100.C-0248, adding 11 epochs between 18 Jan 2018 and 5 Feb 2018. Combined with the VMC epochs these data sets provide 15 and 24 epochs in $J$ and $K_{\mathrm{s}}$ respectively, covering a baseline of $\sim 6$\,years. For our complementary observations the exposure times per pawprint were 90\,s in the $J$ band and 480\,s in the $K_{\mathrm{s}}$ band. The longer exposure time for $K_{\mathrm{s}}$ aims at increasing the likelihood of detecting young variables since several Galactic studies have shown that variability in this filter is common amongst young stars \citep[e.g.][]{Contreras2014, Lucas2017}. 
It also allows us to better probe high-extinction regions, where young stars are preferentially located. 

The pawprint images were reduced and calibrated with the VDFS pipeline v1.5, and  pawprint PSF photometry was performed. Any possible systematic effects owing to different calibration versions for the two datasets are within the photometric errors and hence too small to affect the variability analysis. For this dataset, sources at around $J\approx 19.3$\,mag and $K_{\mathrm{s}}\approx 19.1$\,mag are typically detected in half of the pawprints they are located in.
Further details can be found in Tables \ref{tab:Kepochs_data} and \ref{tab:Jepochs_data}.

\section{Multi-epoch catalogue}
  \label{sec:source_tracking}
  \subsection{Constructing the multi-epoch catalogue}
  \label{subsec:multiepoch_construct}
  We use the PSF catalogue for the deep tile image of LMC 7\_5 \citep{Zivkov2018}, obtained from combining and homogenising  all individual pawprints for the VMC-epochs \citep{Rubele2012,Rubele2015}. This deep catalogue reaches 50\,\% completeness limits at $J\approx 21.3$\,mag and $K_{\mathrm{s}}\approx 20.6$\,mag. All individual pawprint catalogues are cross-matched to the deep catalogue -- in which every source has a source ID -- using a matching radius of $0.5\arcsec$. 
  Overall, $\sim13\,\%$ ($J$) and $\sim6\,\%$ ($K_{\mathrm{s}}$) of the sources in the pawprint catalogues do not have deep catalogue counterparts. The comparatively high fraction of unmatched $J$ sources is likely caused by the small number of VMC epochs; as a consequence, the deep catalogue is not much deeper than the individual pawprint catalogues in $J$.
  This cross-matching removes spurious detections in the individual pawprint catalogues.
    
  For the variability analysis, we further exclude all sources in the pawprint catalogues with $K_{\mathrm{s}}<12.6$\,mag and $J<13$\,mag. This is to avoid saturation and residual non-linearity effects which increase the photometric scatter at bright magnitudes substantially (Fig.\,\ref{fig:Magerr_PP6}).

\subsection{Correcting for residual magnitude offsets}
  \label{subsec:offset_correct}
\begin{figure}
  \centering  
  \includegraphics[width=\columnwidth]{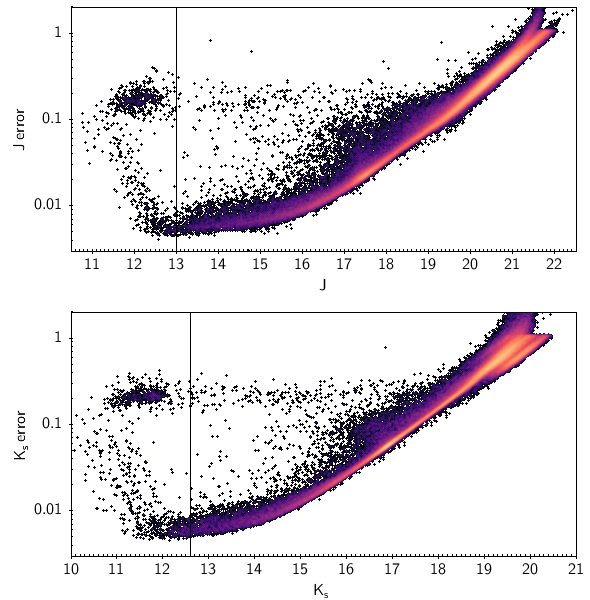}
    \caption{Mean photometric errors versus flux-based mean magnitudes for the $J$ band (top panel) and the $K_{\mathrm{s}}$ band (bottom panel) for pawprint \#6 over all epochs.
    The vertical lines indicate the bright magnitude cutoff
    (Sect.\,\ref{subsec:multiepoch_construct}).}
    \label{fig:Magerr_PP6}
\end{figure}

  The sensitivity of the VISTA system is dependent on the observing conditions. Therefore magnitude zeropoints show variations over long timescales. We corrected for any residual systematic magnitude offsets for all pawprints and detectors. This is done by calculating the difference between the mean magnitudes for the deep catalogue and the pawprint catalogues separately for every detector. The magnitudes in the pawprint catalogues are then shifted accordingly. The applied shifts are generally small, $\sim$99\,\% are within $\pm0.02$ and $\pm0.04\,$mag for the $J$ and $K_{\mathrm{s}}$ bands, respectively. Figure\,\ref{fig:Magerr_PP6} shows for pawprint \#6 the flux-based mean of the corrected magnitudes over all epochs versus the mean of the associated photometric errors (as provided by the PSF pawprint catalogues).
  
\section{Identifying variable star candidates using reduced \texorpdfstring{$\chi^2$}{Chi2}-analysis}
\label{sec:chi2_test}
Different methods are available to identify variables in multi-epoch data. Amongst them, variability detection techniques characterising the light curve scatter performed well in data sets with a small number ($\lesssim 50$) of epochs \citep[see][for an overview]{Sokolovsky2017}. One of these techniques is the reduced $\chi^2$-analysis, which calculates the variance of a series of measurements in a given band and normalises it by the estimated photometric error. For a set of $N$ measurements it is given by
\begin{equation}
    \label{eq:chi2}
    \chi^2 = \frac{1}{N-1}\sum_{i=1}^N \frac{(\mathrm{mag}_i-\overline{\mathrm{mag}})^2}{\sigma_i^2},
\end{equation}
where $\sigma_i$ and $\mathrm{mag}_i$ are the photometric error and the magnitude of the $i$-th measurement, respectively, while $\overline{\mathrm{mag}}$ is the mean magnitude. We considered using the Stetson index \citep{Stetson1996}, which identifies correlated variability between two or more bands using contemporaneous multi-epoch observations. However, only two VMC epochs are observed back-to-back in $J$ and $K_{\mathrm{s}}$. As a consequence, only two of the 13 $K_{\mathrm{s}}$ epochs observed between 2012--2017 could have been used to compute the Stetson index. Hence, we focused on reduced $\chi^2$-analysis for this work.

Pawprint magnitudes and errors are used in Eq.\,\ref{eq:chi2}. However, we treat every pawprint separately so that a $\chi^2$ distribution for each pawprint is obtained. This results in two or more $\chi^2$ values for the majority of sources, since most of a VISTA tile \citep[except at two edges;][]{Sutherland2015} is covered by at least two pawprints. Our approach has the advantage of avoiding any issues related to differences between detectors, since the $\mathrm{mag}_i$ values compared are always for the same detector \citep[for spatial systematics see][]{Gonzales2018}.

\begin{figure*}
	\includegraphics[width=\linewidth]{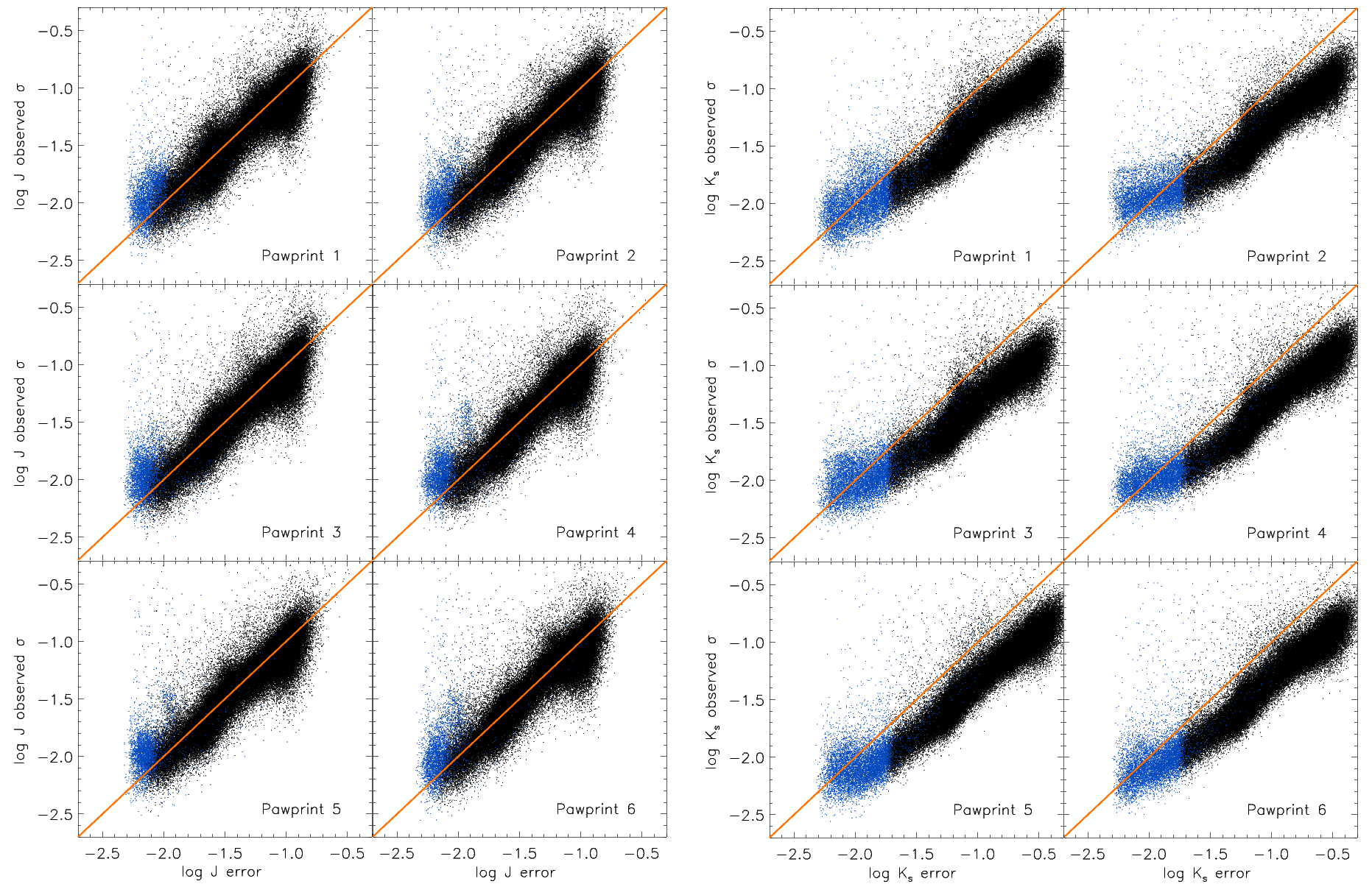}
    \caption{Comparison between the mean photometric errors and the observed standard deviation of the $J$-band (left) and $K_{\mathrm{s}}$-band magnitudes (right), for sources with a minimum of six and ten epochs in $J$ and $K_{\mathrm{s}}$, respectively. Stars with mean magnitudes $J$\,<\,15.3\,mag and $K_{\mathrm{s}}$\,<\,15.5\,mag (i.e. the bright samples) are indicated in blue (see Sect.\,\ref{sec:chi2_test} and Table \ref{tab:chi2_thresholds} for details). The orange line represents the 1:1 relation.}
    \label{fig:pawprints_RMS}
\end{figure*}

In Fig.\,\ref{fig:pawprints_RMS} we compare the mean photometric error with the standard deviation for each light curve in the $J$ and $K_{\mathrm{s}}$ bands, respectively. The distributions should follow a 1:1 relationship (orange line), in which case the photometric errors correctly represent the observed fluctuations. In reality, there is an offset with respect to this relation for the $K_{\mathrm{s}}$ band, which indicates an overestimation of the photometric errors in the pawprint catalogues. This shifts the peak of the $\chi^2$ distribution towards lower values (see also Fig.\,\ref{fig:chi2_dis}). The peak in the $\chi^2$ distributions are populated by non-variable objects which usually represent the majority of the sources. This offset is not observed in the $J$ band $\chi^2$ distributions, i.e. the photometric errors correctly represent the photometric variations.

The $K_{\mathrm{s}}$ distributions (Fig.\,\ref{fig:pawprints_RMS})
approach a noise-floor for small photometric errors, i.e. for bright stars. A noise-floor was also observed in \citep{Rebull2014} based on Spitzer data, but without the systematic offset described above for the $K_{\mathrm{s}}$ data. The differences between pawprints are most pronounced in this regime: the noise-floor in pawprint \#4 is $\approx 0.01$\,mag, but for instance in pawprint \#6 it seems absent. This noise-floor sets in for $K_{\mathrm{s}}\lesssim 15.5$\,mag and $J\lesssim 15.3$\,mag.

As a result we investigated separately the $\chi^2$ distributions for each pawprint, and separated the samples with mean magnitudes brighter and fainter than the values mentioned above. This results in $6\times 2$ distributions per filter, with independently determined $\chi^2$ variability thresholds (Table\,\ref{tab:chi2_thresholds}). These are determined by approximating the logarithmic $\chi^2$ distributions with a Gaussian function and taking its $3\sigma$ value towards the high value tail. The tail is created by objects with magnitude fluctuations exceeding the photometric error, and thus any source beyond the $\chi^2$ threshold is identified as a candidate variable. In general, the $\chi^2$ thresholds in the $K_{\mathrm{s}}$ band are lower than their $J$ counterparts. This is a consequence of the overestimation of the $K_{\mathrm{s}}$ photometric error (Fig.\,\ref{fig:pawprints_RMS}). The bright sample thresholds vary much more than their counterparts for the fainter sample, reflecting the comparatively large pawprint differences in the noise-floor (Fig.\,\ref{fig:pawprints_RMS}). Figures\,\ref{fig:chi2_dis} and \ref{fig:Jchi2_dis} show the logarithmic $\chi^2$ distributions and the Gaussian approximations.

\begin{table}
	\centering
	\caption{3$\sigma$ $\chi^2$ thresholds used to identify photometric variability for each pawprint, filter and magnitude range (bright sample: $J$\,<\,15.3\,mag and $K_{\mathrm{s}}$\,<\,15.5\,mag; faint sample: $J$\,$\geqslant$15.3\,mag and $K_{\mathrm{s}}$\,$\geqslant$\,15.5\,mag).
	  }
	\label{tab:chi2_thresholds}
	\begin{tabular}{lcccc} 
		\hline
		Pawprint & \multicolumn{2}{c}{Bright sample} & \multicolumn{2}{c}{Faint sample}\\
		& $J$ & $K_{\mathrm{s}}$ & $J$ & $K_{\mathrm{s}}$\\
		\hline
		\#1 & 30.0 & 7.38 & 3.97 & 0.86\\
		\#2 & 24.5 & 8.00 & 3.98 & 0.87\\
		\#3 & 28.2 & 5.98 & 4.00 & 0.79\\
		\#4 & 29.1 & 6.00 & 4.04 & 0.73\\
		\#5 & 19.8 & 3.93 & 3.92 & 0.66\\
		\#6 & 28.5 & 3.78 & 4.24 & 0.76\\
		\hline
	\end{tabular}
\end{table}

We require that a source is detected at least in ten and six epochs, respectively for $K_{\mathrm{s}}$ and $J$. Including sources with fewer observations would create a shoulder in the $\chi^2$ distribution towards lower values. These requirements eliminate $\sim 99$\% of sources with $J\gtrsim 19.5$\,mag and $K_{\mathrm{s}} \gtrsim 19$\,mag. However, they offer the best compromise between depth and sensitivity to variations. The final sample for the reduced $\chi^2$-analysis contains 328\,985 sources ($J$ band) and 276\,204 sources ($K_{\mathrm{s}}$ band). There is a large overlap between these samples, thus in total the $\chi^2$ values of 362\,425 stars were analysed.

For a star to be considered a candidate variable its $\chi^2$ value must be above the $3\sigma$ threshold for two pawprints for a given filter. This approach is conservative and it excludes the tile-areas covered by a single pawprint (two stripes of width $5.52^{\prime}$ at the left and right tile-edges in Fig.\ref{fig:LMC75}), but we prioritise reliability over completeness. It also reduces the number of spurious variability detections. In total, 3817 of the 362\,425 stars meet the requirements, 2492 of them in the $J$ band, 2521 in the $K_{\mathrm{s}}$ band and 1196 in both filters.

The spatial distribution of the 3817 candidate variables is shown in Fig.\,\ref{fig:radec_vartot} (top panel). Noticeable are some compact ``clumps'', one of which is at the edge of the N\,44 complex. These are artefacts coincident with very bright sources. Two thin horizontal stripes in the top-right corner, coincident with the top edge of detector \#4, also show an overdensity. These sources are flagged as low reliability objects and removed from the analysis. This results in a final list of 3062 variable star candidates (Table \ref{tab:allvar}), henceforth referred to as the variable star sample.

\section{Analysis of the variable stars}
\label{sec:analysis_varstars}
\subsection{General properties}
\label{subsec:analysis_varstars_prop}

\begin{figure}
  \centering
  \includegraphics[width=\columnwidth]{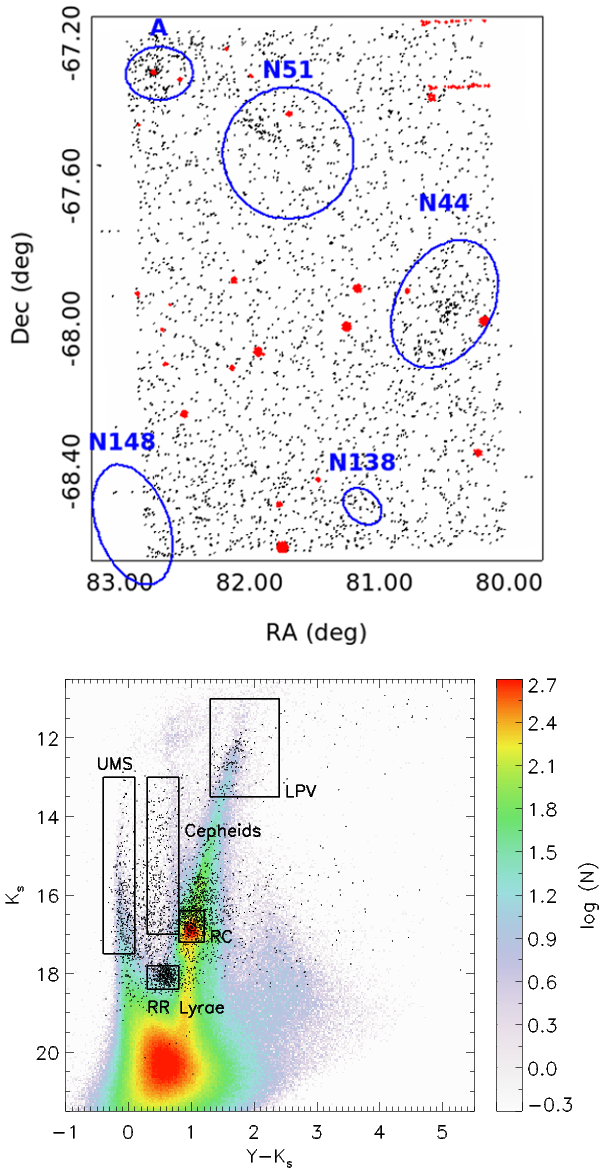}
    \caption{\textbf{Top panel:} Spatial distribution of the 3062 variable star candidates (black points), and of the 755 sources flagged and removed from further analysis (red points). The blue ellipses identify known star forming complexes \citep[see ][]{Zivkov2018}. \textbf{Bottom panel:} Distribution of the 3062 unflagged high reliability variables in a $Y-K_{\mathrm{s}}$ CMD. The background is a Hess diagram of the total stellar population from the deep catalogue. The labelled boxes identify specific types of stars.}
    \label{fig:radec_vartot}
\end{figure}

In some areas in N\,44 and N\,51 a clear increase in the number density of variables is visible (Fig.\,\ref{fig:radec_vartot}, top panel). This could indicate a higher fraction of variables, although the number of stars in general is also increased. Overall, most variables are distributed across the tile similarly to the non-variable field population. This strongly implies that most of the identified variables belong to the field or generally more evolved stellar populations. We note however that there are comparatively more red variables ($Y-K_{\mathrm{s}} \geq 2$\,mag) spatially associated with the N44 and N51 star forming complexes compared to the wider field.

The bottom panel of Fig.\,\ref{fig:radec_vartot} shows the colour$-$magnitude diagram (CMD) of the 3062 variables using photometric data from the deep catalogue. Objects with $K_{\mathrm{s}}>19$\,mag very rarely appear variable; owing to their photometric uncertainties only variables with large amplitudes can be identified. Furthermore, these stars do not meet the minimum number of detections requirement.

\begin{table}
	\centering
	\caption{Number of stars ($N_{\mathrm{tot}}$), number of variable star candidates ($N_{\mathrm{var}}$) and the corresponding fraction for the CMD regions labelled in Fig.\,\ref{fig:radec_vartot} (bottom panel).
	}
	\label{tab:var_fraction}
	\begin{tabular}{lccc} 
		\hline
		Region & $N_{\mathrm{tot}}$ & $N_{\mathrm{var}}$ & fraction (\%)\\
		\hline
		Cepheids     & 3\,631   & 346    & 9.5\\
		RR Lyrae & 4\,787   & 391    & 8.2\\
		UMS      & 4\,572   & 301    & 6.6\\
		LPV      & 4\,292   & 142    & 3.3\\
		RC       & 71\,867  & 364    & 0.5\\
		total    & 362\,425 & 3\,062 & 0.8\\
		\hline
	\end{tabular}
\end{table}

The most prominent features in the variable star CMD are the Red Clump (RC) and the RR Lyrae regions (Fig.\,\ref{fig:radec_vartot}, bottom). A concentration in the RC is unexpected (these stars are not expected to be variable). However the fraction of variable star candidates is in fact considerably smaller than in any other labelled region in the CMD (Table \ref{tab:var_fraction}). Therefore, the apparent enhancement of variables in the RC region is an artifact caused by the high concentration of stars in this CMD region. Extinction could also shift some Cepheids into the RC region. In addition, more metal-rich Cepheids might also contaminate the RC, since they tend to get intrinsically redder with increasing metallicity \citep[e.g.][]{Tammann2003}.

The highest fraction of variables is found in the RR Lyrae and Cepheid regions (see Table \ref{tab:var_fraction}). RR Lyrae are evolved, low mass stars \citep[age $>$10\,Gyr, $M<1\,\mathrm{M_{\odot}}$;][]{Muraveva2018} populating the instability strip at the horizontal branch level. They exhibit radial pulsations similar to Cepheids, which are also evolved stars but more massive \citep[$M_{\mathrm{ini}} \sim 3 - 12\,\mathrm{M_{\odot}}$, e.g.][]{Anderson2016}. In the upper main-sequence (UMS) region the high fraction of variables is likely caused by slowly pulsating B-stars \citep{Wu2018}, $\beta$\,Cephei stars \citep{Stankov2005}, or eclipsing binaries \citep{Kourniotis2014}. The long-period variables (LPVs) are cool giant and supergiant stars with a wide range of periods from 10\,days up to a few years \citep[e.g.][]{Feast1989, Soszynski2009}. Finally, we note that some variables with $1\lesssim Y-K_{\mathrm{s}}\lesssim 3\,$mag could also be quasars \citep{Cioni2013, Ivanov2016}.

\begin{figure}
  \centering
  \includegraphics[width=\columnwidth]{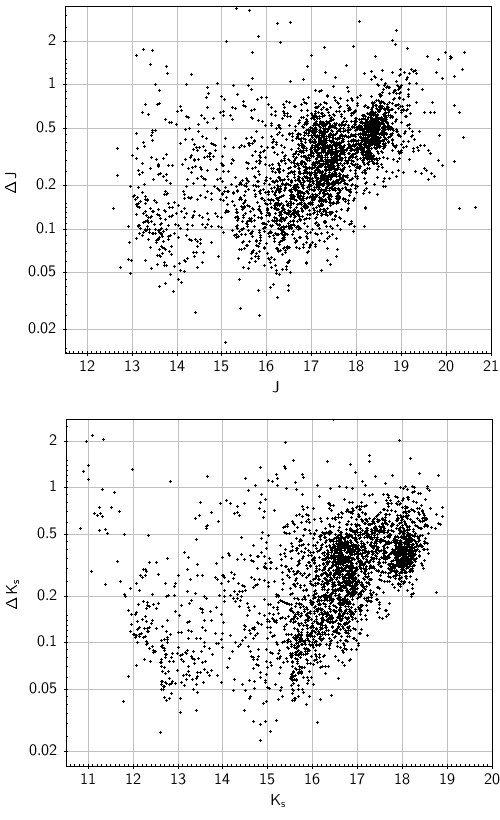}
    \caption{$\Delta J$ vs. $J$ (top) and $\Delta K_{\mathrm{s}}$ vs                 $K_{\mathrm{s}}$ (bottom) for the 3062 variable candidates.
     A few variables are brighter than the bright magnitude cutoffs ($K_{\mathrm{s}}<12.6$\,mag and $J<13$\,mag); these are identified using data from the other band. }
    
    \label{fig:vartot_magvsamp}
\end{figure}
Figure\,\ref{fig:vartot_magvsamp} shows magnitude vs. amplitude plots for the $J$ and $K_{\mathrm{s}}$ bands for the 3062 variables. Amplitudes are defined as the difference between the brightest and the faintest observed magnitude. The minimum amplitude for the variable sample is $\sim$0.03--0.05\,mag; this value is reached for $J\lesssim 17\,$mag and $K_{\mathrm{s}}\lesssim 16.5\,$mag, and is set by the minimum photometric uncertainties. For fainter magnitudes the minimum identifiable amplitudes increase because of the larger photometric errors.

\subsection{OGLE variable counterparts in the VMC survey}
\label{subsec:OGLE comparison}
The increased variability fraction in CMD regions typical of variables is encouraging. However, it is important to establish how effective our approach is, i.e. what fraction amongst a sample of known variables is identified by our method. The Optical Gravitational Lensing Experiment (OGLE) is a sky survey whose original goal was to search for dark matter by detecting microlensing phenomena \citep{Udalski1997}. Since its inception in 1992 it has been invaluable in detecting and classifying a multitude of variable stars \citep[e.g. ][]{Udalski1997b,Soszynski2009,Soszynski2016}.

We used data from the OGLE-III Online Catalog of Variable Stars\footnote{http://ogledb.astrouw.edu.pl/~ogle/CVS/} \citep{Udalski2008}. The OGLE-IV catalogue \citep{Udalski2015} does not provide a list of LPVs, which are essential to extend the magnitude and period range of our variability tests (see below). The OGLE-III survey area covers $\sim 60$\,\% of tile LMC 7\_5, and 3225 OGLE stars have a counterpart (matching radius: $0.5\arcsec$) in the deep VMC catalogue. We removed OGLE sources with VMC counterparts that do not meet our criteria regarding the minimum number of detections (Sect.\,\ref{sec:chi2_test}), or which are brighter than the bright magnitude cut-offs (Sect.\,\ref{subsec:multiepoch_construct}). This left 2276 OGLE sources: 1591 classified as LPVs, 642 RR\,Lyrae  and 43 Cepheids. 

\begin{figure}
  \centering
  \includegraphics[width=\columnwidth]{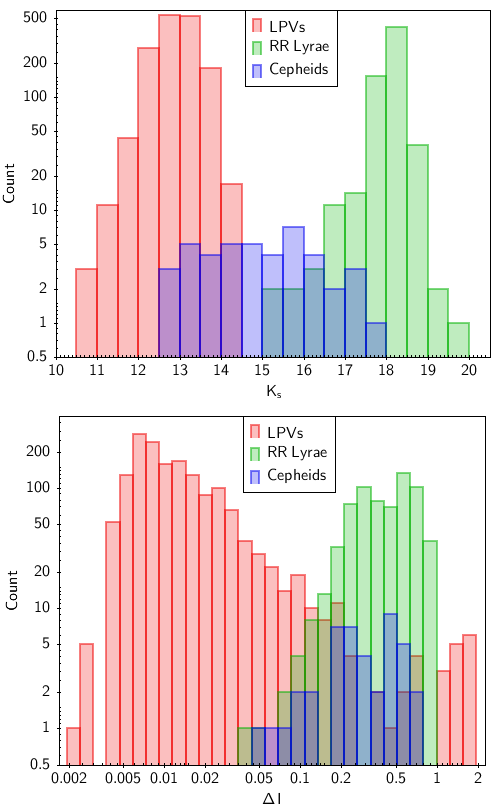}
    \caption{\textbf{Top panel:} Distribution of the $K_{\mathrm{s}}$ magnitudes from the deep catalogue for the 2276 OGLE variables, colour-coded by source type. \textbf{Bottom panel:} Distribution of amplitudes in the $I$-band as given in the OGLE-III catalogue.}
    \label{fig:OGLE_magamp}
\end{figure} 

These three types of variables are distinct in magnitude and amplitude as shown in Fig.\,\ref{fig:OGLE_magamp}. While displaying a wide range of amplitudes, the bright LPVs ($K_{\mathrm{s}}\lesssim 14\,$mag) tend to have small amplitudes ($\Delta I<0.05$\,mag); this subsample is useful to set the lower amplitude limit of our method. Only 109 out of the 1591 LPVs ($\sim 7$\%) were identified by us as variable. However, the fraction varies strongly among the different LPV sub-types. We identified all the 14 Mira variables and $\sim 66$\% of the semi-regular variables (SRVs; 41 of 62).
Approximately 55\,\% of the LPVs with $\Delta I>0.05$\,mag were identified as variable; this fraction rises to $\sim 66$\,\% and $\sim 90$\,\% for $\Delta I>0.1$\,mag and $\Delta I>0.2$\,mag, respectively. In the $J$-band $\sim$\,50\% of all LPVs with $\Delta J \geq 0.08$\,mag were identified as variable; in the $K_{\mathrm{s}}$-band the same fraction is achieved for $\Delta K_{\mathrm{s}}\geq 0.11$\,mag.

The amplitudes of the 43 Cepheids are largely above the amplitude limits established in the LPV analysis ($\Delta J \gtrsim 0.1$\,mag and $\Delta K_{\mathrm{s}} \gtrsim 0.1$\,mag). Due to their relatively high luminosities, this led to a high success rate in variable identification ($\sim 80$\,\%, 34 of 43). Similarly $\gtrsim 90\,$\% of the 642 RR\,Lyrae exhibit amplitudes above 0.1\,mag in both filters. However, as a result of their comparatively low brightnesses we identified only 302 RR\,Lyrae ($\sim 47$\,\%) as variable.
\citet{Moretti2016} examined the recovery rate of OGLE variables for the VMC data using the  VISTA Science Archive \citep[VSA;][]{Cross2012} variability flag. Those rates are broadly consistent for Cepheids, LPV-Mira and LPV-SRVs. We report a significant improvement for RR Lyrae (47\% vs. 3\%). This might be related to the larger number of $J$ epochs in our combined data (Sect.\,\ref{subsec:VMC_data} and \ref{subsec:complementary_data}), since most RR Lyrae are identified as variable in the $J$ band. However, our recovery rate is worse for LPVs classified as OGLE Small Amplitude Red Giants (4\% vs. 16\%), which probably results from our requirement of significant variability in at least two pawprints.

\begin{figure}
  \centering
  \includegraphics[width=\columnwidth]{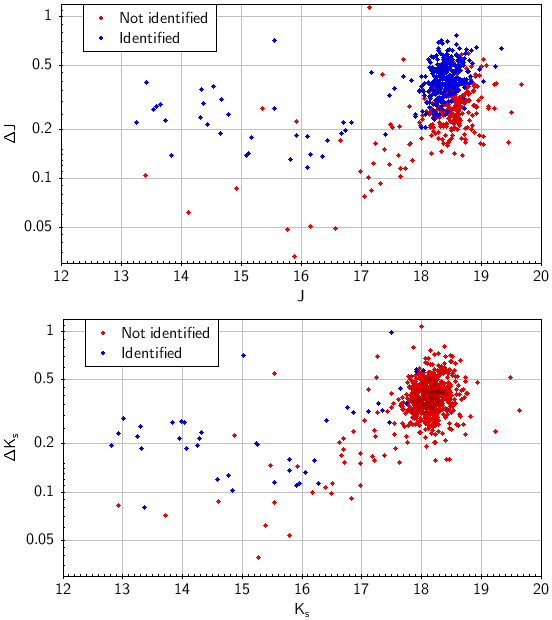}
    \caption{$\Delta J$ vs. $J$ (top) and $\Delta K_{\mathrm{s}}$ vs. $K_{\mathrm{s}}$ (bottom) for the Cepheids and RR\,Lyrae, identified as variable (blue), or not identified as variable (red), in each band.}
    \label{fig:OGLE_RRCep}
\end{figure}

Given their similar amplitudes, both the Cepheid and RR\,Lyrae samples can be used to constrain the typical amplitude needed for variability identification across a wide range of magnitudes. Figure \ref{fig:OGLE_RRCep} shows amplitude versus magnitude for these samples. As expected, the identified variables tend to have larger amplitudes. For $J\lesssim 17$\,mag most sources with $\Delta J > 0.1\,$mag are identified as variable; by $J\approx 18\,$mag this limit is $\Delta J \approx 0.2\,$mag and it reaches $\Delta J \approx 0.5\,$mag for $J\approx 19\,$mag. For $K_{\mathrm{s}} \lesssim 16$\,mag the minimum required amplitude is $\Delta K_{\mathrm{s}} \approx 0.1\,$mag, increasing to $\Delta K_{\mathrm{s}} \approx 0.5$\,mag for $K_{\mathrm{s}} \approx 18$\,mag.

Based on this analysis of the OGLE counterparts in the VMC data we conclude the following. To reliably detect variability, i.e. with a completeness of $\gtrsim 50$\,\%, an amplitude of at least $\sim 0.1$\,mag is needed for stars with $J\lesssim 17$\,mag and $K_{\mathrm{s}} \lesssim 16$\,mag. At fainter magnitudes the required amplitude is at least $\sim 0.5$\,mag for  $J\approx 19$\,mag and $K_{\mathrm{s}} \approx 18$\,mag. Hence, we cannot investigate variability in the low and intermediate mass PMS populations identified by \citet{Zivkov2018}, which have typical magnitudes $K_{\mathrm{s}}\sim 18 - 22\,$mag, and so we restrict our analysis to the light curves of massive YSOs in tile LMC 7\_5 (see Sect.\,\ref{sec:YSO_sample}).

\subsection{Periodicity tests}
\label{subsec:period_tests}
The VMC survey was not designed with the goal of determining periods of variable stars. Nevertheless it is useful to consider how well we can recover periods using our combined data set, since periodicity can provide additional clues about the origin of the observed variability. We made use of periods listed in the OGLE catalogue.

\begin{figure}
  \centering
  \includegraphics[width=\columnwidth]{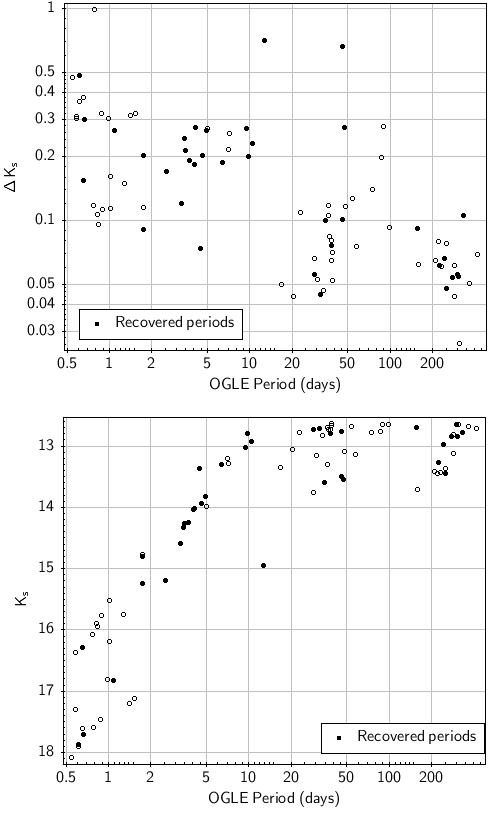}
    \caption{\textbf{Top panel:} $\Delta K_{\mathrm{s}}$ vs. OGLE period for the 87 OGLE stars found to be variable in the $K_{\mathrm{s}}$-band. The recovered periods are highlighted. \textbf{Bottom panel:} $K_{\mathrm{s}}$ vs. OGLE period for the same sample.}
    \label{fig:periodsteps}
\end{figure}

Since the number of epochs is larger we focused on the $K_{\mathrm{s}}$ band data for this analysis; in total, 87 OGLE variables were identified as variables: 34 Cepheids, 45 LPVs and 8 RR Lyrae. This sample covers a wide range of periods and amplitudes (Fig.\,\ref{fig:periodsteps}). We note that intrinsic stellar properties affect our analysis. For instance LPVs are bright with low-amplitude variability in most cases, while Cepheids tend to be fainter, have shorter periods, but larger amplitudes.

We used the \textsc{LombScargle} class from the astropy library\footnote{http://people.bolyai.elte.hu/$\sim $ sic/astropy/stats/lombscargle.html}. It is designed to detect periodic signals in unevenly spaced observations \citep{Lomb1976,Scargle1982}. As input parameters it uses the dates of observation, magnitudes, photometric errors and list of periods to be tested. For each required period, the routine calculates a Fourier power and a corresponding false alarm probability (FAP). The period with the highest power is assumed to be the best-fitting period of the light curve, usually under the condition that the FAP must be below a certain value.

Different lists of periods containing between 100 and 2000 periods in a logarithmic spacing between $P_{\mathrm{min}}\approx 0.55\,$d and $P_{\mathrm{max}}=1000\,$d were tested by evaluating how well the calculated periods matched the OGLE periods. Based on the distribution of the deviations between calculated and OGLE periods, a period was deemed recovered if this difference was within $\pm 40\,$\% of the OGLE period. This rather large tolerance is acceptable given the cadence, time span and number of epochs in the data, and the fact that we make no a-priori assumptions about the shape of the light curves. We found that a list of 500 periods provides the best compromise: above 500 periods there is no further improvement in the recovery rate; below 500 periods the resolution is too coarse.

Overall, only 36 out of 87 OGLE periods were successfully recovered (19 Cepheids, 15 LPVs and two RR Lyrae). For Cepheids we have therefore a similar period recovery rate ($\sim 56$\%) as \citet{Moretti2016}, but note that we allow for a larger tolerance. In the range 2$-$20\,d for 15 out of 19 stars we determined the correct periods ($\sim$\,79\%). For $P>20\,$d the success rate dropped markedly (15 out of 44, $\sim$\,34\% periods identified correctly). As can be seen in the top panel of Fig.\,\ref{fig:periodsteps}, the amplitudes are not responsible for this behaviour. Instead the high cadence observations of our 2018 dataset probably increased the sensitivity preferentially in the 2$-$20\,d range. For the shortest periods ($<2\,$\,d) only 6 out of 24 periods (25\%) were recovered. This low success-rate may be partially due the objects' fainter magnitudes, even though magnitudes already decrease gradually for sources with $P\lesssim 5$\,d (Fig.\,\ref{fig:periodsteps}, bottom panel). More importantly, the observing cadence is not suited to reliably identify such short periods: with the exception of two epochs, all other epochs have at least 1\,day separation.

Comparing the FAP distributions for the entire OGLE sample and for the stars with successfully recovered periods did not suggest a reasonable FAP threshold that could have been used as an indicator for a correctly determined period. Therefore, our observational parameters (number of epochs, total time baseline, cadence) do not seem suited for an automated periodicity search. For the subsequent analysis we did rely on a FAP threshold, but only considered a light curve as periodic if a visual inspection corroborated the maximum power period.

The findings of our periodicity analysis can be summarised as follows. Periods shorter than 2\,d were rarely recovered irrespective of amplitude. The best sensitivity to periodicity was achieved in the 2$-$20\,d range, where $\sim$\,79\% of the OGLE periods were recovered. For $P>20\,$d period recovery was also considerably reduced. Regarding periodic YSOs we would thus expect to be most sensitive to variability caused by rotational modulation of stellar spots, or by phenomena related to the inner disc \citep{Wolk2013}.

\section{The YSO sample}
\label{sec:YSO_sample}
To investigate variability characteristics of young stars in tile LMC 7\_5, we assemble a reliable sample of massive YSOs from the literature, and analyse the corresponding subsample of variables.

\subsection{\textit{Spitzer} source selection}
\label{subsec:Spitzer_sample}
Several wide-field studies on massive star formation in the LMC used \textit{Spitzer} data for the identification and analysis of YSOs \citep[e.g.][]{Whitney2008, Gruendl2009, Carlson2012}. These YSOs
 tend to be in an early evolutionary stage (mostly Class I) in which stars are more frequently variable \citep[e.g. ][]{Cody2014} and high amplitude variability is more common \citep[e.g. ][]{Contreras2014}. These YSOs are also luminous enough to be reliably detected in the individual pawprints. To compile the massive YSO sample ($\gtrsim 3\,\mathrm{M_{\odot}}$; \citealt{Carlson2012}) we select stars from the following studies:

\begin{itemize}
  \item \textbf{\citet{Gruendl2009}:} This study carried out independent aperture photometry on the images from the \textit{Spitzer} SAGE programme \citep{Meixner2006}. Applying colour and magnitude cuts to remove evolved stars and background galaxies, they identified a sample of 2910 potential YSOs. About 150 of these are located within tile LMC\,7\_5 in areas covered by at least two pawprints. To further reduce the likelihood of contaminantion by field stars we selected only sources which are spatially associated with regions containing PMS candidates \citep{Zivkov2018}, which leaves 79 YSO candidates.

  \item \textbf{\citet{Carlson2012}:} This work focused on nine large star forming complexes in the LMC, inclusing N\,44 and N\,51 in tile LMC\,7\_5. This work uses a combination of SAGE IRAC and MIPS data (3.6$-$24\,$\mathrm{\mu m}$), $UBVI$ photometry from the Magellanic Cloud Photometric Survey \citep[MCPS;][]{Zaritsky2002,Zaritsky2004} and $JHK_{\mathrm{s}}$ data from the InfraRed Survey Facility (IRSF; \citealt{Kato2007}). A selection based on \textit{Spitzer} colours and magnitudes was applied, followed by spectral energy distribution (SED) fitting using YSO models from \citet{Robitaille2006}. This resulted in a sample of 1045 well-fit YSO candidates out of which 242 are located in N\,44 and N\,51. Based on the SED fitting 157 stars were classified as Class I, 64 as Class II and 21 as Class III.

  \item \textbf{Spectroscopic samples:} Based on the analysis of \citet{Gruendl2009}, 294 highly embedded objects were selected by \citet{Seale2009} for follow-up \textit{Spitzer} Infrared Spectrograph \citep[IRS;][]{Houck2004} spectroscopy. In total, 277 stars had spectral features consistent with embedded YSOs. Of these, 42 YSOs are located in tile LMC\,7\_5 and were added to the YSO sample without further spatial constraints, since this set is considered highly reliable. In addition we also added five spectroscopically confirmed YSOs from \citet{Jones2017}, which brings the number of high-reliability YSOs to 47.
\end{itemize}

Taking into account the overlap among the three samples, we compiled a sample of 345 unique massive YSOs.
\subsection{VMC massive YSO counterparts}

\subsubsection{VMC catalogue matching}
\label{subsec:matching_Spitzer_VISTA}

\begin{figure}
  \centering
  \includegraphics[width=\linewidth]{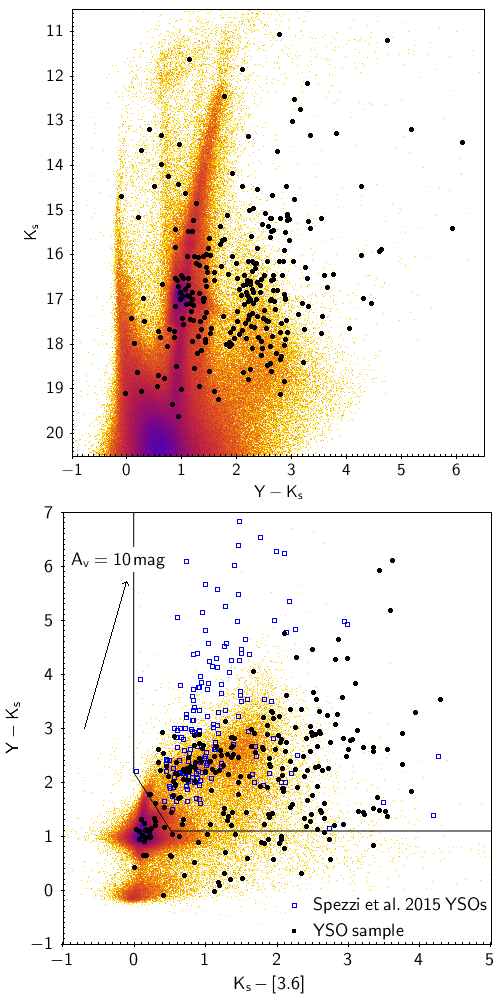}
    \caption{\textbf{Top panel:} CMD of VMC counterparts in the \textit{Spitzer} YSO sample. The total stellar population within the tile is displayed in the background. \textbf{Bottom panel:} $Y-K_{\mathrm{s}}$ vs. $K_{\mathrm{s}}-[3.6]$ CCD showing the Galactic YSOs from \protect\citet{Spezzi2015} (open blue squares) and massive LMC YSOs (solid black circles). Solid lines show the colour cuts and the arrow represents the interstellar reddening vector for $A_{V}=10\,$mag \citep{Nishiyama2009}.}
    \label{fig:YSOmatches}
\end{figure}

The 345 YSOs were matched with the deep VMC catalogue. This links every \textit{Spitzer} YSO with the source ID provided by the deep catalogue, which allows the tracking of the YSOs throughout all epochs and pawprints (see Sect.\,\ref{sec:source_tracking}). We select a matching radius of $0.75^{\prime\prime}$, which provides VMC counterparts to 305 YSOs. The $Y-K_{\mathrm{s}}$ CMD for these sources is shown in Fig.\,\ref{fig:YSOmatches} (top panel). Their distribution indicates some contamination, in particular at the RC locus. A coincidental match test suggests a $\sim$\,15\% probability for a random match. Reducing the matching radius is however problematic due to the astrometric uncertainty of \textit{Spitzer}\footnote{https://irsa.ipac.caltech.edu/data/SPITZER/docs/irac/\\iracinstrumenthandbook/30/}.

\subsubsection{Additional colour criteria}
\label{subsubsec:colour_cuts}
\citet{Spezzi2015} combined VISTA and \textit{Spitzer} observations to investigate the young stellar content of the Lynds\,1630 star forming region located in the Orion molecular cloud. Using multi-colour criteria, 186 YSO candidates were selected with ages of 1--2\,Myr (Spezzi-YSOs). Based on their YSO sample we devised criteria for the removal of contaminants. The Spezzi-YSOs have typically lower masses, however no significant colour differences across the mass range are expected and our colour cuts (see below) are very generous.

Figure\,\ref{fig:YSOmatches} (bottom panel) displays a ($K_{\mathrm{s}}-[3.6]$, $Y-K_{\mathrm{s}}$) colour$-$colour diagram (CCD), including both the Galactic and LMC samples. Both samples deviate substantially from the underlying distribution of more evolved stars, but also from each other. Whilst some sources from the Spezzi-YSOs are very red with $K_{\mathrm{s}}-[3.6] > 2\,$mag, most show rather moderate colours. In contrast, the massive LMC YSOs are predominantly redder in $K_{\mathrm{s}}-[3.6]$. The Spezzi-YSOs are mostly Class II objects (126 out of 186, \citealt{Spezzi2015}), indicating that these stars have prominent circumstellar discs. On the other hand, the LMC sample is expected to consist mostly of Class I objects \citep[e.g.][]{Carlson2012}, which are still surrounded by massive envelopes. Hence, the difference in the CCD distributions is likely caused by a difference in evolutionary stage. 

We used empirical colour-cuts (solid lines in Fig.\,\ref{fig:YSOmatches}, bottom panel) as follows:
$Y-K_{\mathrm{s}} \geqslant 1.1$\,mag,
$K_{\mathrm{s}}-[3.6] \geqslant 0$\,mag and 
$Y-K_{\mathrm{s}} \geqslant 2.2 - 2.2\times(K_{\mathrm{s}}-[3.6])$.
YSOs with colours outside of these boundaries were removed from further analysis. This eliminates the enhancement of sources located in the RC region at $Y-K_{\mathrm{s}} \approx 1.1$\,mag and $K_{\mathrm{s}}-[3.6] \approx 0.15$\,mag, and left 207 YSOs for subsequent visual inspection (Sect.\,\ref{subsubsec:vis_exam}).

\subsubsection{Visual examination}
\label{subsubsec:vis_exam}
Colour-cuts alone cannot identify background galaxies. In addition, some \textit{Spitzer} sources are associated with multiple VMC counterparts. We made use of the higher VISTA-resolution to identify such sources. The visual examination was performed on colour composite VISTA images ($YJK_{\mathrm{s}}$ bands); the appearance of the 207 YSO candidates was judged based on their shape and intensity profile.

\begin{figure}
  \centering
  \includegraphics[width=\columnwidth]{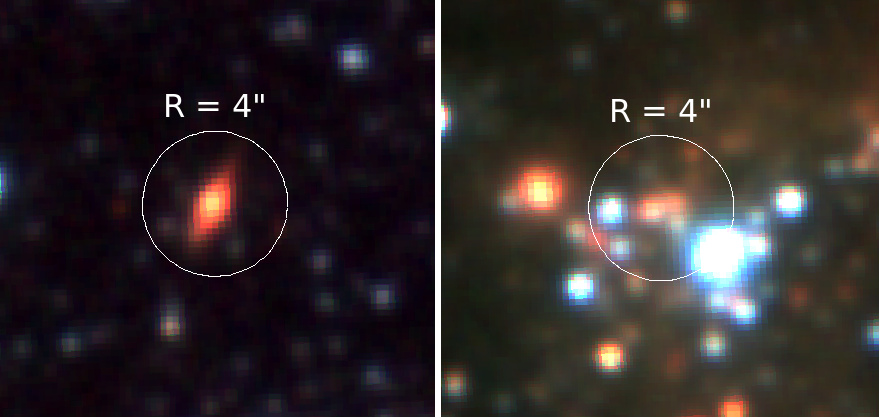}
    \caption{VISTA RGB composites with $Y$ in blue, $J$ in green and $K_{\mathrm{s}}$ in the red showing a likely background galaxy seen edge-on (left) and a YSO resolved into multiple VMC sources (right). The radius of the circle corresponds to 1\,pc at the LMC distance of $50\pm 2\,$kpc \citep{DeGrijs2014}.}
    \label{fig:YSOvisual_examples}
\end{figure}

We identified 12 spatially extended sources that are likely galaxies (respectively seven and five from the \citealt{Carlson2012} and \citealt{Gruendl2009} samples). Figure\,\ref{fig:YSOvisual_examples} (left panel) shows one such example. These sources were removed from the sample. In 29 cases, \textit{Spitzer} YSOs are resolved into two or more sources in the VMC images (Fig.\,\ref{fig:YSOvisual_examples} right panel). In this example, the $K_{\mathrm{s}}$ band brightness is dominated by two red sources that are both likely to be young. The VMC-counterpart is flagged, but not removed from the YSO sample.

\begin{figure}
  \centering
  \includegraphics[width=\columnwidth]{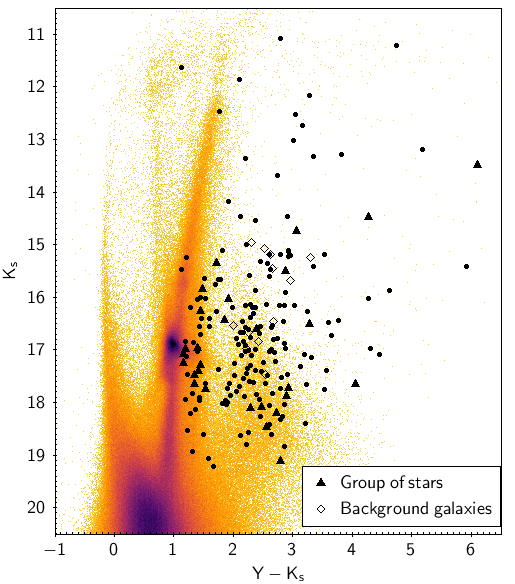}
    \caption{CMD showing the YSO sources with VMC counterparts that meet the colour criteria (Sect.\,\ref{subsubsec:colour_cuts}). Sources identified as groups of YSOs and possible background galaxies are identified. The latter are removed from the YSO sample. The background displays the total stellar population.}
    \label{fig:CMD_backgals}
\end{figure}

The CMD of the sources that meet the colour criteria in Sect.\,\ref{subsubsec:colour_cuts} are shown in Fig.\,\ref{fig:CMD_backgals}. The position of the spatially extended sources is indeed consistent with the expected locations of background galaxies \citep{Kerber2009}; by contrast the sources found to be part of a group are not confined to a tight range in colour. 

After removing the background galaxies, we were left with 195 highly reliable YSO candidates. We checked if the YSOs met the requirements for the reduced $\chi^2$-analysis (Sect.\,\ref{sec:chi2_test}) and if they were below our bright magnitude cutoff (Sect.\,\ref{subsec:multiepoch_construct}) in at least one band. This removed 22 sources (21 did not have enough detections, one was too bright in both bands), so that the final YSO sample contains 173 stars.
They have a median magnitude of $\overline{K_{\mathrm{s}}} \approx 16.7\,$mag with 10th and 90th percentiles of $K_{\mathrm{s, 10}}\approx 14.5\,$mag and $K_{\mathrm{s, 90}}\approx 17.8\,$mag, respectively. For the $J$ band the corresponding values are $\overline{J} \approx 18.3\,$mag, $J_{10}\approx 16.3\,$mag and $J_{90}\approx 19.5\,$mag. 

\section{Results and discussion}
\label{sec:YSO_varresults}
Our sample of 173 YSOs was examined for variability according to the selection criteria from Sect.\,\ref{sec:chi2_test}.  As a result 39 YSOs were found to be variable; their light curves are shown in the appendix (Fig.\,\ref{fig:lc_appendix}). For each object the $K_{\mathrm{s}}$ light curves were examined and classified based on their appearance. Light curve classifications are presented in Sect.\,\ref{subsec:LC_classes}; the properties of variable YSOs (e.g. periodicity, amplitude, etc.) are discussed in Sect.\,\ref{subsec:YSO_var_prop}. A detailed listing of the properties of all 39 YSO variables can be found in Table \ref{tab:ysovar}.

\subsection{Light curve classification}
\label{subsec:LC_classes}
We classified the light curves based on their shape, which may be related to the physical processes responsible for the variability. The classes follow the scheme adopted by \citet{Contreras2017}. While they focus on high amplitude stars ($\Delta K_{\mathrm{s}} > 1\,$mag), this scheme was also used by \citet{Teixeira2018} for stars of lower amplitudes. The classes are eruptives, dippers, faders, short-term variables (STV) and long-period variable YSOs (LPV-YSO).

For the classification we focus mainly on the $K_{\mathrm{s}}$ light curves; the number of epochs is larger and therefore the time-sampling is better. All YSOs are brighter in the $K_{\mathrm{s}}$-band which implies smaller photometric errors compared with the $J$ measurements. As will be shown in Sec.\,\ref{subsec:YSO_var_prop}, most YSOs are identified as variables only in the $K_{\mathrm{s}}$-band. However, in some cases the $J$ light curve is helpful in constraining the possible origin of the variability. In the next subsections we describe each classification in detail. 

\subsubsection{Eruptive}
\label{subsubsec:eruptive}
Eruptives are aperiodic YSOs which experience outbursts resulting in an increase in luminosity. The outbursts are typically of long duration (>\,1\,yr, \citealt{Contreras2017}), but some YSOs display shorter outbursts. This type of light curve is thought to be evidence of accretion events, or of changes in the line-of-sight extinction in which case the star would move along the reddening vector in CCDs \citep{Contreras2014}.

\begin{figure}
  \centering
  \includegraphics[width=\columnwidth]{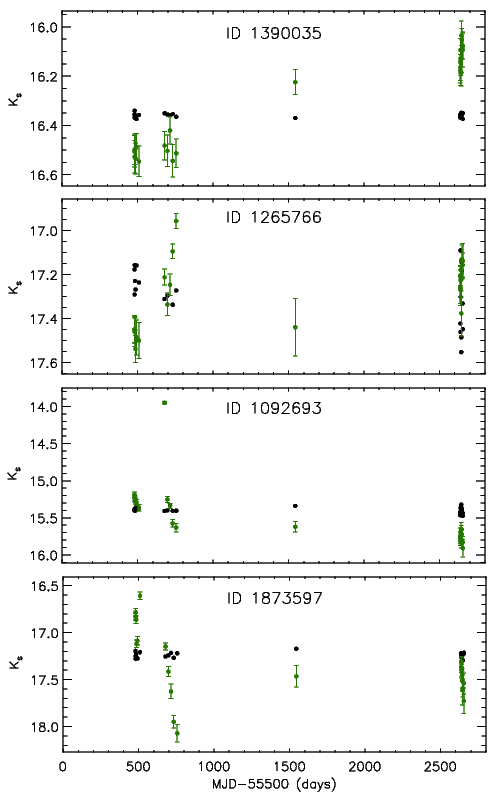}
    \caption{Examples of light curves of stars identified as eruptives (green circles). The black circles show the light curve of a nearby non-variable star (separation $< 3^{\prime}$) with similar $K_{\mathrm{s}}$-band magnitude for comparison. Since the photometric errors are very similar we have omitted plotting the error bars for the companion star.}
    \label{fig:eruptives}
\end{figure}

The YSOs classified as eruptives display a variety of light curves; four distinct examples are shown in Fig.\,\ref{fig:eruptives}. The top panel shows the light curve of a YSO with a slow increase in $K_{\mathrm{s}}$ band magnitude over several years, the most common type observed in our data. Such light curves are typical of FU Orionis (FUor)-type sources \citep{Herbig1977}, however known Galactic examples usually exhibit larger amplitudes and a steeper increase in brightness (e.g. \citealt{Hartmann1996}), although slower rises were also observed \citep{Contreras2017b}. The $K_{\mathrm{s}}$ band magnitude may increase beyond our observational window, so the observed amplitude ($\Delta K_s \approx 0.5\,$mag for this particular example) is likely a lower limit. The second panel shows a YSO with an outburst, after which the brightness returns to its quiescent level before a second outburst occurs. This object could be periodic and the Lomb-Scargle analysis calculates a period of $\sim 970$\,days. In the third panel the object shows a single-epoch significant outburst ($\Delta K_{\mathrm{s}} \approx 1.5\,$mag), typical of EX Lupi (EXor)-type sources \citep{Herbig2007,Moody2017}. This light curve is unique in our sample. Finally, the star in the bottom panel experiences a sudden magnitude increase ($\Delta K_{\mathrm{s}}\approx 0.7\,$mag) followed by a significant drop ($\Delta K_{\mathrm{s}} \approx 1.5\,$mag), before returning to its apparent quiescent brightness. Because of its behaviour after the initial outburst, we classified this source also as a dipper. Based on its light curve a combination of processes, like an accretion event followed by some kind of obscuration, seems likely.

In total, 12 YSOs were classified as eruptive variables which makes this the most common class in our sample. In six cases the light curves are similar to that shown in the top panel, thus allowing us to calculate only a lower limit for the duration of the outburst ($\sim 4$\,yr in the example shown). All three YSO variables with $\Delta K_{\mathrm{s}} > 0.6\,$mag (see Fig.\,\ref{fig:YSOamplitudes} bottom panel) are eruptives, although one of them (Fig.\,\ref{fig:eruptives}, bottom panel) is also classified as a dipper.

\subsubsection{Fader}
\label{subsubsec:fader}
Faders are aperiodic variables with declining luminosity. This decline can be slow over the course of months or years or relatively sudden. The physical origins of the change in brightness are similar to those of the eruptives. They can either be a star returning to quiescent levels after an outburst, or the fading can be caused by a long lasting increase in line-of-sight extinction \citep{Findeisen2013}. If extinction is the cause of the fading, then it would be more pronounced in the $J$-band.

\begin{figure}
  \centering
  \includegraphics[width=\columnwidth]{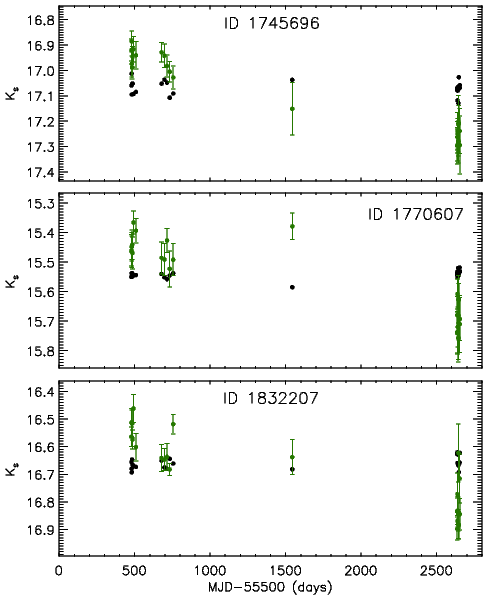}
    \caption{As Fig.\ref{fig:eruptives} but for stars classified as faders.}
    \label{fig:faders}
\end{figure}

The faders show relatively similar light curves (examples in Fig.\,\ref{fig:faders}). They all exhibit a gradual magnitude decrease, for the duration of the observing window (top panel), or starting at some later point during the observations (middle panel). In some cases there is short term variability superimposed onto the long-term dimming (bottom panel). This could hint at a combination of multiple physical processes, e.g. additional modulation caused by star spots.

Overall, 10 YSOs were classified as faders. In two cases there is very little or no fading in the $J$-band magnitudes, making an obscuration event very unlikely. One object experienced a significantly larger magnitude drop in the $J$-band ($\Delta J \approx 0.55\,$mag, $\Delta K_{\mathrm{s}} \approx 0.3\,$mag), strongly supporting extinction as the cause of the fading. For all other faders extinction might play only a minor role, as the fading in the $J$-band is slightly less pronounced than in the $K_{\mathrm{s}}$ band (Sec.\,\ref{subsec:YSO_longvar}).

\subsubsection{Dipper}
\label{subsubsec:dipper}
Dippers experience fading followed by a return to their normal brightness. This class is generally associated with extinction events and they share some light curve properties with the faders. 

\begin{figure}
  \centering
  \includegraphics[width=\columnwidth]{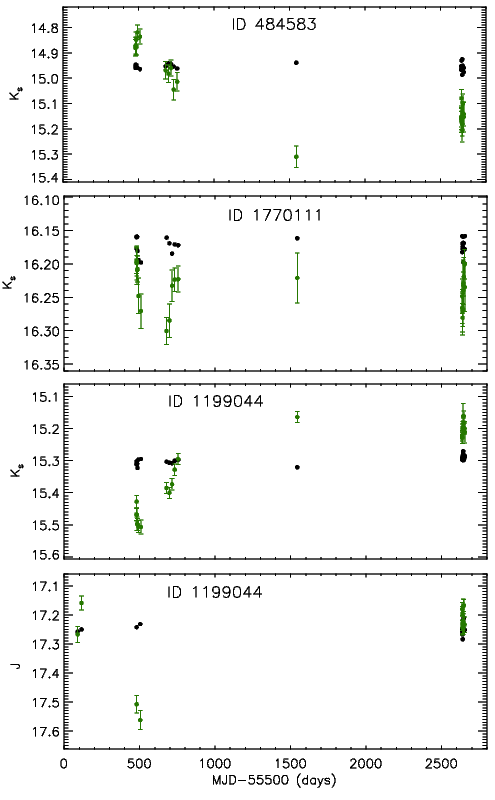}
    \caption{As Fig.\ref{fig:eruptives} but for stars classified as dippers. The light curve in the bottom panel is from the same object as that in the panel above it, but for the $J$ band.}
    \label{fig:dippers}
\end{figure}

Some of the dippers had not yet returned to their presumably normal magnitudes by the end of the observation window (Fig.\,\ref{fig:dippers}, top panel); the brightness remains lower than in the early epochs. The second panel shows an object with a relatively shallow dip ($\Delta K_{\mathrm{s}} \approx 0.1\,$mag) of $\sim 250$\,days duration; there are four YSOs with this type of light curve. Another example is shown in the bottom two panels. The $K_{\mathrm{s}}$ light curve has a narrow dip but the beginning of the dimming was missed. The first $J$ band epochs were observed almost a year before any $K_{\mathrm{s}}$ epochs and indicate that the object indeed returned to a quiescent magnitude. The source shown in Fig.\,\ref{fig:eruptives} (bottom panel) is classified as a dipper as well as an eruptive. In total, we classified seven YSOs as dippers.

\subsubsection{Short-term variable}
\label{subsubsec:stv}
This class comprises stars which show either periodic or aperiodic variations in their luminosity over timescales of $<100$\,days. This type of variability can be explained either by photospheric phenomena modulated by the rotation of the star \citep{Wolk2013}, by orbital variations in the disc extinction \citep{Rice2015}, or by variable accretion \citep{Bouvier2003}. Two STVs have a best-fitting period in the 2$-$20\,d range (3.5\,d and 2.1\,d), which is where the OGLE analysis shows a relatively high success rate in recovering known periods (Sec.\,\ref{subsec:period_tests}). The fact that only two such sources are found is broadly consistent with our YSO sample being in an early evolutionary stage. Periods in this range are often caused by rotational modulation, which is more readily observed in more evolved Class II or III objects \citep{Contreras2017}.

\begin{figure}
  \centering
  \includegraphics[width=\columnwidth]{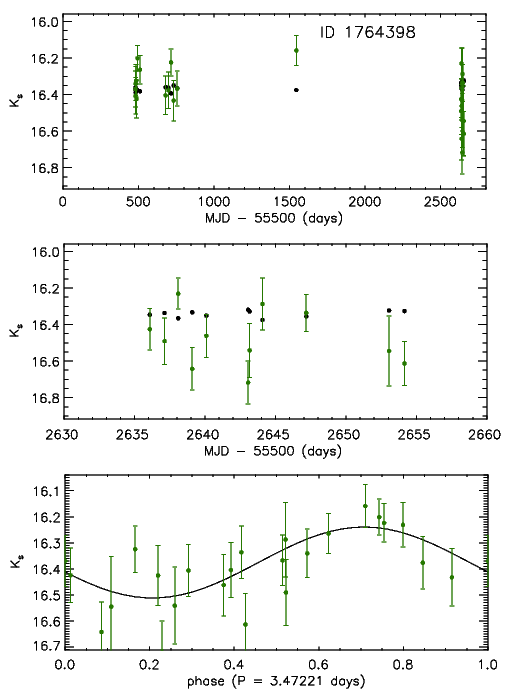}
    \caption{\textbf{Top panel: } $K_{\mathrm{s}}$ light curves of a YSO classified as STV. \textbf{Middle panel: } Zoom in on the high cadence observations at the end of the observation window. \textbf{Bottom panel: } Folded light curve of the same STV, with a sinusoidal model ($P\approx 3.5\,$d) overplotted.}
    \label{fig:STVs}
\end{figure}

One of the periodic STVs is shown in Fig.\,\ref{fig:STVs}. The top panel, covering the full monitoring period, does not display long-term luminosity trends. Zooming in on the high cadence epochs (middle panel), short-timescale variations dominate, which are considerably larger than the variations seen in the light curves of the comparison star. The bottom panel shows the folded light curve with a sinusoidal model ($P \approx 3.5\,$d) overplotted. The model matches the measurements, but the amplitude is rather small and the folded light curve shows considerable scatter. Hence, our claim of periodicity is tentative.

Eight YSOs were classified as STVs. They mostly display small amplitudes ($\Delta K_{\mathrm{s}} < 0.3$\,mag). Since the $J$ light curves show similar amplitudes (when available), changes in extinction seem unlikely. More likely are variations associated either with stellar spots or moderate changes in the accretion rate.

\subsubsection{Long-period variable YSO}
\label{subsubsec:lpv_yso}
This class of YSO variables consists of stars exhibiting periodic variability with  periods $>100$\,days. Such light curves could arise from variable accretion modulated by a binary companion \citep{Hodapp2012}, or by obscuration due to a circumbinary disc \citep{Contreras2017}. The inspection of the light curves lends support to periodicity in three cases only, with calculated periods of 451, 494 and 618 days.

\begin{figure}
  \centering
  \includegraphics[width=\columnwidth]{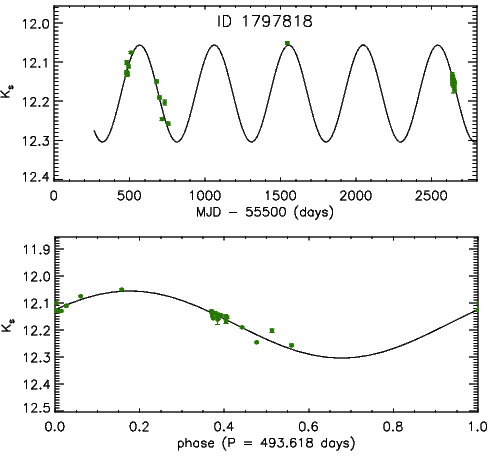}
    \caption{$K_{\mathrm{s}}$ light curve of a YSO classified as LPV-YSO (top, $P\approx 494\,$d) and the corresponding folded light curve (bottom). A sinusoidal fit at the estimated period is overplotted.}
    \label{fig:LPVs_period}
\end{figure} 

One of these sources (shown in Fig.\,\ref{fig:LPVs_period}) is a YSO identified as variable in the $J$ band only, since its $K_{\mathrm{s}}$ band magnitude is brighter than the cutoff limit. Based on the $J$ light curve, this YSO was classified as eruptive. However, the $K_{\mathrm{s}}$ measurements can be fitted by a $\sim 494$\,d period, hinting that this YSO might indeed be periodic. Consequently, we classified this source also as a potential LPV-YSO. Given that the sensitivity to long periods is low (see Sec.\,\ref{subsec:period_tests}), the classification of all three LPV-YSOs is tentative. In summary, five YSOs were tentatively classified as periodic: two with $2<P<20\,$d (STVs, Sec.\,\ref{subsubsec:stv}) and three with $P>100\,$d (LPV-YSOs).

\subsubsection{Unclassified}
\label{subsubsec:no_class}
\begin{figure}
  \centering
  \includegraphics[width=\columnwidth]{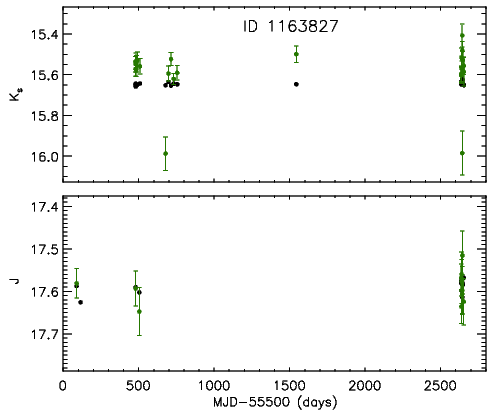}
    \caption{$K_{\mathrm{s}}$ and $J$ light curves of the unclassified YSO variable.}
    \label{fig:non_class}
\end{figure}

One YSO defies classification into any of the classes above. This object is from the \citet{Gruendl2009} sample and its light curves are shown in Fig. \,\ref{fig:non_class}. Two prominent and short-duration dips of similar amplitude were detected in the $K_{\mathrm{s}}$-band (top panel) suggesting an eclipsing binary. The $J$ light curve (bottom panel) does not show these features: the second dip should have been visible in the $J$-band, since back-to-back observations in both filters were obtained. In the OGLE-IV database of eclipsing binaries this star is not listed and a visual inspection of the images did not reveal obvious artifacts. This is the only object without a light curve classification (Tables \ref{tab:amp_YSOvars} and \ref{tab:class_YSOvars}).

\subsection{Properties of the variable YSOs}
\label{subsec:YSO_var_prop}
\begin{figure}
  \centering
  \includegraphics[width=\columnwidth]{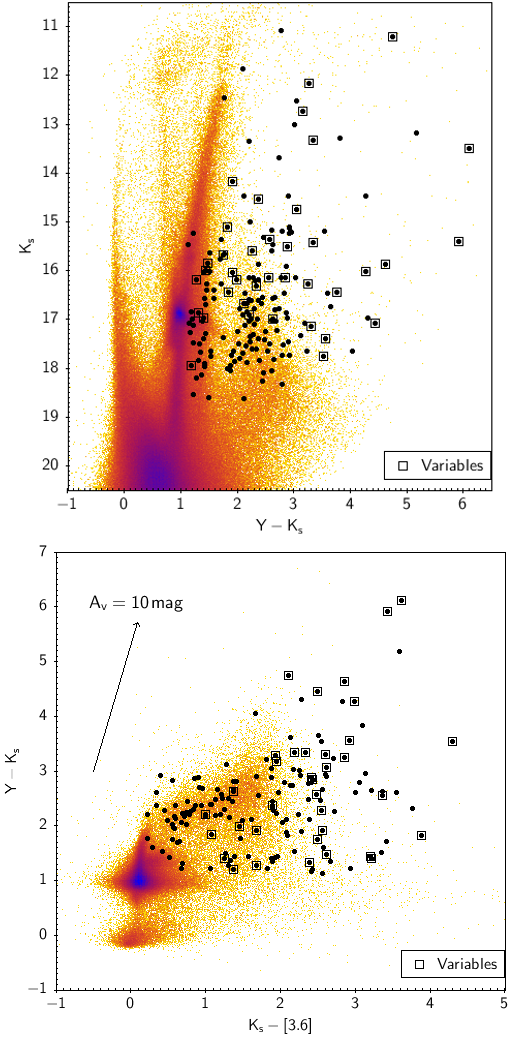}
    \caption{CMD (top) and CCD (bottom) with the YSO sample (circles) and the variable YSOs (squares) identified.}
    \label{fig:CMD_YSOvariables}
\end{figure}

Figure\,\ref{fig:CMD_YSOvariables} shows the location of the YSO variables in a CMD (top panel) and a CCD (bottom panel). Variability seems more prevalent amongst the brighter YSOs, but this is likely a selection effect (larger amplitudes required for fainter sources). The overall variability fraction is $\sim23$\,\% (39 of 173), increasing for $K_{\mathrm{s}} < 17$\,mag to $\sim 33$\,\% (34 of 104). For YSOs with $K_{\mathrm{s}} < 16$\,mag (the regime where we achieve the best sensitivity, see Sec.\,\ref{subsec:OGLE comparison}) the fraction is $\sim 37$\,\% (18 of 49).

NIR-variability is more common amongst very red sources. Slightly over half of the YSOs with $Y-K_{\mathrm{s}}>3$\,mag (16 of 30) show variability, compared with only $\sim 16\,$\% (23 of 143) for $Y-K_{\mathrm{s}}<3$\,mag. The mean colours for the entire sample of 173 YSOs are $\langle Y-K_{\mathrm{s}} \rangle = 2.43\pm 0.06\,$mag and $\langle K_{\mathrm{s}}-[3.6] \rangle = 1.88\pm 0.10\,$mag. For the YSO variables the corresponding colours are $\langle Y-K_{\mathrm{s}} \rangle _{\mathrm{var}} = 2.82\pm 0.07\,$mag and $\langle K_{\mathrm{s}}-[3.6] \rangle_{\mathrm{var}} = 2.46\pm 0.08\,$mag. The values for the YSO variables cannot be explained by higher reddening alone. Assuming that the redder stars are least evolved, we see a trend towards more widespread variability for less evolved sources. However, a selection effect could be at play since redder objects tend to be brighter in the $K_{\mathrm{s}}$-band, making a variable identification more likely.

Of the 39 variable YSOs, 31 were identified as variable only in the $K_{\mathrm{s}}$ band, while six met the criteria in both bands. Two YSO variables were identified based on $J$ band data, since both objects are brighter than the cutoff limit in the $K_{\mathrm{s}}$ band. Upon investigation, both showed $K_{\mathrm{s}}$ variations that were larger than expected from the photometric errors alone. The distinct efficiency in variability detection between the two bands is likely caused by the fact that the YSOs are considerably fainter in $J$. Of the 28 YSOs with $13<J<17$\,mag, 5 are variable in the $J$ band ($\sim 18\,$\%). For the $K_{\mathrm{s}}$ band there are 31 variables out of 97 YSOs ($\sim 32\,$\%) for $13<K_{\mathrm{s}}<17$\,mag. Assuming that the variability fraction in the $J$ band is the same as that in the $K_{\mathrm{s}}$ band, one would expect on average $\approx 9$ variables in a sample of 28 YSOs.
Using Poissonian statistics the probability of having five or fewer variables is $P(k\leq 5) \approx 0.122$, which shows that a lower observed fraction is not unlikely. In addition, the smaller number of epochs in the $J$ band reduces the sensitivity to certain types of variables.

\begin{landscape}
\begin{table}
	\centering
	\caption{Properties of the 39 identified YSO variables. From left to right listed are the star catalogue IDs, coordinates, $YJK_{\mathrm{s}}$ magnitudes taken from the deep catalogue, band(s) in which the star was identified as variable, amplitudes in $J$ and $K_{\mathrm{s}}$, light curve classification, period (if applicable) and reference which identifies the star as a likely YSO.}
	\label{tab:ysovar}
	\begin{tabular}{rccccccccccl} 
		\hline
		ID & RA\,(J2000) & Dec\,(J2000) & $Y$ & $J$ & $K_{\mathrm{s}}$ & variable band(s) & $\Delta J$ & $\Delta K_{\mathrm{s}}$ & light curve class & period & reference \\
		& & & (mag) & (mag) & (mag) & & (mag) & (mag) & & (days) & \\
		\hline
		293446  & $81\fdg5220$ & $-68\fdg6026$ & 15.91 & 15.08 & 12.74 & $K_{\mathrm{s}}$      & 0.25 & 0.24 & dipper           & - & \citet{Jones2017}\\
		335048  & $82\fdg7261$ & $-68\fdg5746$ & 19.60 & -     & 13.49 & $K_{\mathrm{s}}$      & -    & 0.57 & fader            & - &  \citet{Seale2009}\\
		446107  & $81\fdg0556$ & $-68\fdg4996$ & 15.95 & 14.48 & 11.20 & $J$                   & 0.25 & 0.51 & eruptive         & - &  \citet{Jones2017}\\
		451859  & $81\fdg0590$ & $-68\fdg4958$ & 18.70 & 17.92 & 16.14 & $K_{\mathrm{s}}$      & 0.54 & 0.20 & eruptive         & - &  \citet{Gruendl2009}\\
		484583  & $82\fdg4269$ & $-68\fdg4734$ & 17.81 & 17.05 & 14.75 & $K_{\mathrm{s}}$      & 0.40 & 0.49 & dipper           & - & \citet{Gruendl2009}\\
		694596  & $81\fdg2493$ & $-68\fdg3246$ & 17.42 & 16.93 & 15.67 & $K_{\mathrm{s}}$      & 0.07 & 0.32 & eruptive         & - & \citet{Gruendl2009}\\
		1037373 & $80\fdg5704$ & $-68\fdg0677$ & 20.51 & 19.04 & 15.88 & $K_{\mathrm{s}}$      & 0.31 & 0.21 & STV              & - & \citet{Seale2009}\\
		1059694 & $80\fdg6360$ & $-68\fdg0505$ & 18.77 & 17.61 & 15.42 & $K_{\mathrm{s}}$      & 0.40 & 0.46 & STV              & - & \citet{Seale2009}\\
		1092693 & $80\fdg7048$ & $-68\fdg0247$ & 16.94 & 16.45 & 15.11 & $J$, $K_{\mathrm{s}}$ & 0.68 & 1.96 & eruptive         & - & \citet{Seale2009}\\
		1125390 & $80\fdg2549$ & $-67\fdg9995$ & 18.39 & 17.96 & 16.98 & $J$, $K_{\mathrm{s}}$ & 0.40 & 0.29 & eruptive         & - & \citet{Carlson2012}\\
		1138726 & $80\fdg5356$ & $-67\fdg9894$ & 18.38 & 17.98 & 16.98 & $K_{\mathrm{s}}$      & 0.24 & 0.42 & STV              & 2.1 & \citet{Gruendl2009}\\
		1158992 & $80\fdg5306$ & $-67\fdg9741$ & 21.33 & 19.43 & 15.41 & $K_{\mathrm{s}}$      & 0.16 & 0.29 & STV              & - & \citet{Carlson2012}\\
		1163827 & $80\fdg5509$ & $-67\fdg9703$ & 18.38 & 17.60 & 15.50 & $K_{\mathrm{s}}$      & 0.13 & 0.58 & -                & - & \citet{Gruendl2009}\\
		1169275 & $80\fdg5076$ & $-67\fdg9661$ & 21.29 & 21.08 & 17.76 & $K_{\mathrm{s}}$      & -    & 0.58 & eruptive         & - & \citet{Carlson2012}\\
		1199044 & $80\fdg4804$ & $-67\fdg9430$ & 17.94 & 17.25 & 15.37 & $J$, $K_{\mathrm{s}}$ & 0.40 & 0.35 & dipper           & - & \citet{Gruendl2009}\\
		1249325 & $80\fdg6152$ & $-67\fdg9036$ & 20.30 & 18.85 & 16.02 & $K_{\mathrm{s}}$      & 0.24 & 0.24 & STV              & - & \citet{Seale2009}\\
		1265766 & $80\fdg9628$ & $-67\fdg8907$ & 20.95 & 20.41 & 17.39 & $K_{\mathrm{s}}$      & -    & 0.58 & eruptive         & - & \citet{Carlson2012}\\
		1304816 & $80\fdg3720$ & $-67\fdg8599$ & 19.63 & 18.82 & 17.00 & $K_{\mathrm{s}}$      & 0.36 & 0.26 & fader            & - & \citet{Carlson2012}\\
		1390035 & $80\fdg4801$ & $-67\fdg7917$ & 19.52 & 18.60 & 16.28 & $K_{\mathrm{s}}$      & 0.32 & 0.51 & eruptive         & - & \citet{Seale2009}\\
		1395605 & $80\fdg5498$ & $-67\fdg7872$ & 19.01 & 18.26 & 16.15 & $K_{\mathrm{s}}$      & 0.27 & 0.24 & LPV-YSO          & 451 & \citet{Gruendl2009}\\
		1505996 & $81\fdg6235$ & $-67\fdg6957$ & 16.67 & 15.82 & 13.32 & $J$, $K_{\mathrm{s}}$ & 0.56 & 0.32 & fader            & - & \citet{Seale2009}\\
		1580793 & $81\fdg6088$ & $-67\fdg6351$ & 18.80 & 18.28 & 16.68 & $K_{\mathrm{s}}$      & 0.43 & 0.48 & eruptive         & - & \citet{Carlson2012}\\
		1601639 & $81\fdg5354$ & $-67\fdg6181$ & 20.20 & 19.14 & 16.44 & $K_{\mathrm{s}}$      & 0.29 & 0.37 & dipper           & - & \citet{Carlson2012}\\
		1690496 & $81\fdg3582$ & $-67\fdg5451$ & 19.14 & 18.80 & 17.95 & $J$, $K_{\mathrm{s}}$ & 0.59 & 0.47 & fader            & - & \citet{Carlson2012}\\
		1745696 & $81\fdg7214$ & $-67\fdg4993$ & 19.11 & 18.46 & 16.91 & $K_{\mathrm{s}}$      & 0.35 & 0.41 & fader            & - & \citet{Carlson2012}\\
		1746046 & $81\fdg5127$ & $-67\fdg4990$ & 17.96 & 17.29 & 16.04 & $K_{\mathrm{s}}$      & 0.81 & 0.39 & fader            & - & \citet{Carlson2012}\\
		1748223 & $81\fdg3347$ & $-67\fdg4973$ & 18.65 & 17.86 & 16.31 & $K_{\mathrm{s}}$      & 0.15 & 0.19 & LPV-YSO          & 618 & \citet{Carlson2012}\\
		1757648 & $81\fdg3552$ & $-67\fdg4899$ & 16.09 & 15.66 & 14.17 & $J$, $K_{\mathrm{s}}$ & 0.18 & 0.20 & STV              & - & \citet{Gruendl2009}\\
		1764398 & $81\fdg8011$ & $-67\fdg4846$ & 17.44 & 16.99 & 16.00 & $K_{\mathrm{s}}$      & 0.29 & 0.56 & STV              & 3.5 & \citet{Gruendl2009}\\
		1770111 & $82\fdg1423$ & $-67\fdg4801$ & 18.19 & 17.68 & 16.20 & $K_{\mathrm{s}}$      & 0.22 & 0.11 & dipper           & - & \citet{Carlson2012}\\
		1794798 & $81\fdg8983$ & $-67\fdg4603$ & 18.19 & 17.78 & 16.86 & $K_{\mathrm{s}}$      & 0.32 & 0.90 & eruptive         & - & \citet{Carlson2012}\\
		1796271 & $81\fdg9243$ & $-67\fdg4591$ & 17.47 & 17.07 & 16.19 & $K_{\mathrm{s}}$      & 0.10 & 0.17 & fader            & - & \citet{Gruendl2009}\\
		1797818 & $81\fdg9037$ & $-67\fdg4579$ & 15.44 & 14.57 & 12.16 & $J$                   & 0.20 & 0.21 & dipper, LPV-YSO  & 494 & \citet{Seale2009}\\
		1814857 & $81\fdg5179$ & $-67\fdg4441$ & 16.92 & 16.07 & 14.54 & $K_{\mathrm{s}}$      & 0.32 & 0.25 & fader            & - & \citet{Gruendl2009}\\
		1828191 & $81\fdg6072$ & $-67\fdg4333$ & 20.46 & 19.71 & 17.15 & $K_{\mathrm{s}}$      & 0.22 & 0.28 & eruptive         & - & \citet{Carlson2012}\\
		1832207 & $81\fdg2751$ & $-67\fdg4301$ & 18.30 & 17.34 & 16.45 & $K_{\mathrm{s}}$      & 0.13 & 0.43 & fader            & - & \citet{Carlson2012}\\
		1837040 & $81\fdg9832$ & $-67\fdg4263$ & 17.34 & 17.24 & 15.86 & $K_{\mathrm{s}}$      & 0.09 & 0.18 & STV              & - & \citet{Gruendl2009}\\
		1839817 & $82\fdg0079$ & $-67\fdg4241$ & 17.86 & 17.22 & 15.59 & $K_{\mathrm{s}}$      & 0.08 & 0.22 & fader            & - & \citet{Gruendl2009}\\
		1873597 & $82\fdg0320$ & $-67\fdg3971$ & 21.54 & 20.21 & 17.09 & $K_{\mathrm{s}}$      & 0.44 & 1.47 & eruptive, dipper & - & \citet{Carlson2012}\\
		\hline
	\end{tabular}
\end{table}
\end{landscape}

\begin{figure}
  \centering
  \includegraphics[width=\columnwidth]{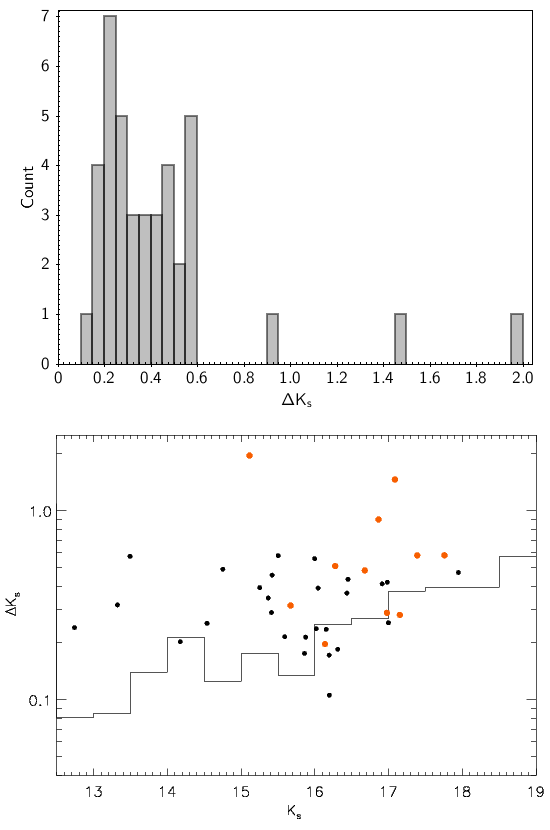}
    \caption{\textbf{Top panel:} Distribution of the $\Delta K_{\mathrm{s}}$ amplitudes for the variable YSOs. \textbf{Bottom panel:} $\Delta K_{\mathrm{s}}$ vs. $K_{\mathrm{s}}$ plot for all variable YSOs (circles) and eruptives (red). The histogram shows the median $\Delta K_{\mathrm{s}}$ in 0.5 magnitude bins based on the total variable sample.}
    \label{fig:YSOamplitudes}
\end{figure}

The YSO amplitudes are mostly in the range $0.1 < \Delta K_{\mathrm{s}} < 0.6\,$mag as shown in Fig.\,\ref{fig:YSOamplitudes} (top panel) -- the observed amplitudes are in fact lower limits. Only two of the LMC YSO variables have $\Delta K_{\mathrm{s}} > 1\,$mag. Nevertheless, YSOs display above average amplitudes compared with the total sample of 3062 variables. Figure\,\ref{fig:YSOamplitudes} (bottom panel) shows $\Delta K_{\mathrm{s}}$ vs. $K_{\mathrm{s}}$ for both samples. The histogram represents the median $\Delta K_{\mathrm{s}}$ for the total variable sample in 0.5\,mag bins. Most YSO variables (30 of 39) are located above the histogram bins, indicating that YSO variability is characterised by larger than average amplitudes. For the Milky Way this is also the case: $\sim 50$\,\% of all variables with $\Delta K_{\mathrm{s}} > 1\,$mag are likely YSOs \citep{Contreras2017}.

Faders and STVs show similar mean amplitudes with a relatively small scatter (Table \ref{tab:amp_YSOvars}), while for the dippers and eruptives a significantly larger scatter is observed. Eruptives display the largest mean amplitude and the five YSOs with the highest amplitude belong to this class. The LPV-YSOs show small amplitudes.

\begin{table}
	\centering
	\caption{Number of variables, mean amplitudes, standard deviations and median amplitudes for all classes. The sum is 41 instead of 39 because one star was classified as eruptive and dipper, and another as eruptive and LPV-YSO.}
	\label{tab:amp_YSOvars}
	\begin{tabular}{lcccc}
		\hline
		 & $N$ & mean($\Delta K_{\mathrm{s}}$) & SD($\Delta K_{\mathrm{s}}$) & median($\Delta K_{\mathrm{s}}$)\\
		 &   & (mag)                        & (mag)			 & (mag) \\
		\hline
		Eruptives & 12 & 0.67 & 0.51 & 0.51 \\
		Faders    & 10 & 0.35 & 0.12 & 0.39 \\
		Dippers   &  7 & 0.46 & 0.43 & 0.35 \\
		STVs      &  8 & 0.32 & 0.13 & 0.29 \\
		LPV-YSO   &  3 & 0.21 & 0.02 & 0.21 \\
		unclassified  &  1 & -- & -- & -- \\
		\hline
	\end{tabular}
\end{table}

\begin{figure}
  \centering
  \includegraphics[width=\columnwidth]{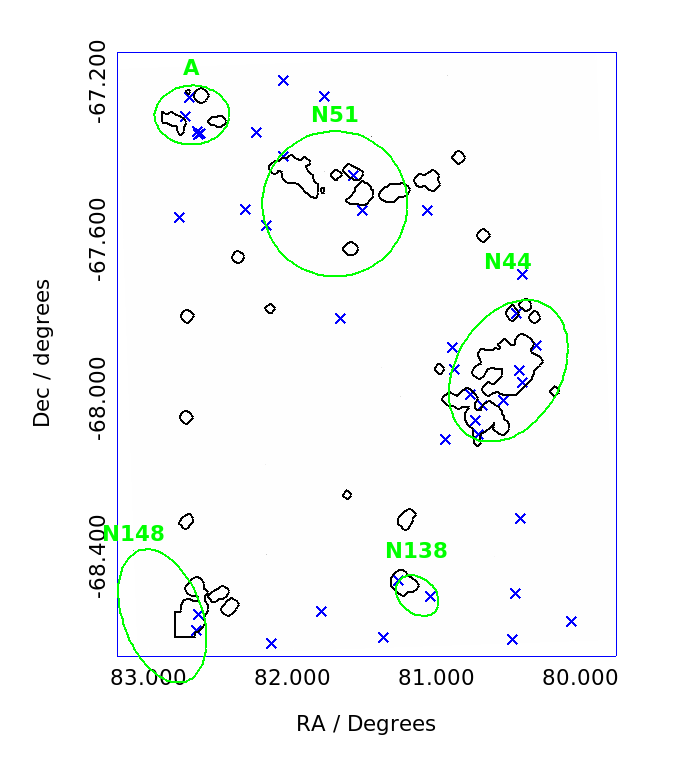}
    \caption{Spatial distribution of the 40 high-amplitude variables ($\Delta K_{\mathrm{s}}>1\,$mag; crosses). The green ellipses show known star forming complexes, the black contours indicate areas with significant PMS populations \citep{Zivkov2018}. There is a tendency for the high-amplitude variables to be associated with sites of ongoing star formation.}
    \label{fig:highamp_vars}
\end{figure}

For ten YSO variables we have IRS-based spectral classifications from \citet{Jones2017}; they are classified as either YSOs or compact H\,{\sc II} regions, which are often very difficult to distinguish at the distance of the LMC. For 19 YSO variables we know the SED class from \citet{Carlson2012}; the majority (13, $\sim 68$\%) are listed as Class I. Note however that $\sim$\,65\% of sources in that sample are Class I.
Therefore there does not seem to be a clear trend between spectral or SED classification and light curve classification or amplitude. 

Of the 2521 stars identified as variable in the $K_{\mathrm{s}}$-band (Sec.\,\ref{sec:chi2_test}), we find 40 variables with $\Delta K_{\mathrm{s}}>1\,$mag. Of these sources, 25 are broadly spatially associated with the PMS structures outlining star forming regions
(see \citealt{Zivkov2018} and Fig.\,\ref{fig:highamp_vars}). Two were included in our sample of 39 YSOs and are classified as eruptives. The remainder either have no previous \textit{Spitzer} classification (12 objects), or no \textit{Spitzer} point-source counterpart (9 objects). An additional two sources have SEDs that are not well-fitted by YSO models according to \citet{Carlson2012}. We checked the SIMBAD database\footnote{http://cds.u-strasbg.fr/} and found that none of the 25 high amplitude variables associated with PMS structures are classified as evolved stars. Therefore, most of these sources may be YSOs but they were not included in our high-reliability YSO sample. Thus we can speculate that large amplitude variables ($\Delta K_{\mathrm{s}}>1\,$mag) could well be primarily associated with YSOs, similar to what is seen in the Galaxy \citep{Contreras2017}.

\subsection{Long-term variability and colour analysis}
\label{subsec:YSO_longvar}
 
We investigated the long-term behaviour of the YSO variables by calculating the difference between the mean magnitudes in two time intervals: February to March 2012 (two $J$ epochs, seven $K_{\mathrm{s}}$ epochs) and January to February 2018 (11 epochs in both filters). Two earlier $J$-epochs from early 2011 are available but they have no counterparts in the $K_{\mathrm{s}}$ band.

\begin{figure}
  \centering
  \includegraphics[width=\columnwidth]{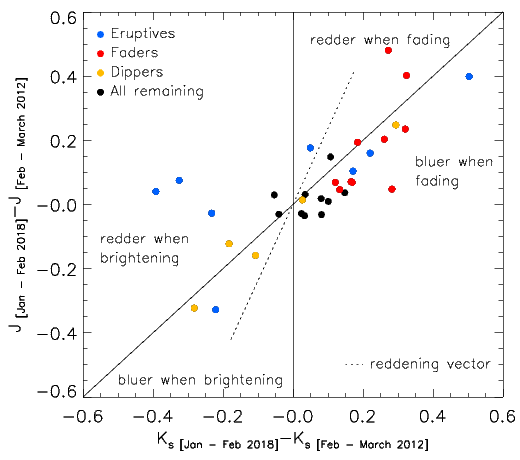}
    \caption{Amplitudes for the $J$ and $K_{\mathrm{s}}$ bands, using the magnitude averages for the time intervals (see text). The circles are colour-coded according to the YSO light curve classification.}
    \label{fig:YSO_longvar}
\end{figure}

Figure\,\ref{fig:YSO_longvar} shows the differences in mean magnitude for these two time intervals. Since most sources are located in the top-right and the bottom-left quadrants, there is a generally positive correlation between the $J$ and the $K_{\mathrm{s}}$ bands. For objects located close to the 1:1 diagonal, the variability is colour-independent. Objects with extinction-related variability should appear close to the reddening vector (dotted line): such variable stars get redder when fading and bluer when brightening. Most YSO variables are concentrated near the 1:1 diagonal, suggesting that extinction is not the main cause of the observed variability. However, grey (i.e.\,wavelength-independent) extinction can be caused by large (compared to the wavelength) dust grains. The presence of grey-opacity dust was modelled by \citet{Menshchikov1999} as resulting from the coagulation of small grains in circumstellar discs. This phenomenon was inferred from observations of massive Galactic stars \citep{Patrriarchi2003}, where it was associated with local cloud effects. In the 30\,Doradus complex in the LMC, a grey-opacity component was invoked to explain the observed slope of the extinction law as a function of wavelength as derived from an analysis of RC stars \citep{DeMarchi2016}.

All but two faders show close to colour-independent variability, hence either extinction is not the main cause of the observed variability or the presence of large grains (grey extinction) is required.
One is close to the reddening line, indicative of variability associated with standard extinction. The other fader becomes considerably bluer when fading. Since three eruptives redden significantly while brightening, this fader could be an eruptive returning to a quiescent state. Such a colour-behaviour is often seen for variables where unsteady accretion \citep{Poppenhaeger2015} or changes in the disc geometry \citep{Rice2015} are the main physical processes driving the observed variations. The other eruptives in our sample exhibit either only small colour changes or are consistent with variability driven by changes in the line-of-sight extinction. Dippers are all distributed near the 1:1 diagonal, showing a negligible colour change.
For the STVs, LPV-YSOs and the unclassified source we generally do not find significant long-term changes in their magnitudes or colours.

\subsection{Comparison with Galactic studies}
\label{subsec:gal_comparison}
Galactic variability studies of massive YSOs in the NIR show a significant spread in variability fraction ranging from values similar to ours ($\sim 26\,$\% by \citealt{Teixeira2018}) to $>50$\,\% \citep{Borissova2016}, although the latter study also included less massive objects. Many non-physical factors influence this fraction, such as time baseline, cadence, sensitivity to small amplitudes, or the criteria for classification as variable. Nevertheless, it is apparent that NIR-variability is a common feature also among YSOs in the LMC.

Table \ref{tab:class_YSOvars} shows how many variables belong to a given class based on two Galactic YSO studies \citep{Contreras2017,Teixeira2018} and this work. Both Galactic studies use NIR data from the VISTA Variables in The V{\'i}a L{\'a}ctea (VVV) public survey with an observing window spanning five years, which is similar to our baseline of $\sim 6$\,years. Eruptive behaviour is seen in roughly one quarter of the YSO samples in all three studies. Noticeable is the large number of STVs in both Galactic studies, compared with our sample. This is likely caused by our relatively small number of epochs compared with the VVV data ($\sim$ 50 epochs in \citealt{Contreras2017}).  The LPV-YSO fraction is also significantly lower in this work as a consequence of the low sensitivity to long periods discussed in Sec.\,\ref{subsec:period_tests}.

\begin{table}
	\centering
	\caption{Breakdown of YSO light curve classifications in this work and from \citet[][CP17]{Contreras2017} and \citet[][T18]{Teixeira2018}.}
	\label{tab:class_YSOvars}
	\begin{tabular}{lccc}
		\hline
		 & CP17 & T18 & This work \\
		\hline
		Eruptives & 106 & 41 & 12 \\
		Faders    & 39  & 18 & 10 \\
		Dippers   & 45  & 20 & 7  \\
		STVs      & 162 & 49 & 8  \\
		LPV-YSOs  & 65  & 62 & 3  \\
		Eclipsing binaries & 24 & 0 & 0 \\
		Unclassified & 0 & 0 & 1 \\
		\hline
		Total count    & 441 & 190& 41 \\
		\hline
	\end{tabular}
\end{table}

In the sample analysed by \citet{Borissova2016} most variables ($\sim$\,78\%) have amplitudes between 0.2\,mag and 0.5\,mag, with the remainder exhibiting larger amplitudes. This is consistent with our analysis: we report $\Delta K_{\mathrm{s}} > 0.5\,$mag for $\sim$\,25\% of our sample (10 of 39 YSOs). In contrast, \citet{Teixeira2018} find that almost 90\% of their YSO variables have $\Delta K_{\mathrm{s}} > 0.5$\,mag. A Kolmogorov-Smirnov test returned a probability that the two samples are drawn from the same distribution (null-hypothesis probability) of $1.3\times 10^{-7}$. We note that 54 of the YSO variables from \citet{Teixeira2018} are associated with clumps detected at $870\,\mathrm{\mu m}$. Hence, they are likely in an early evolutionary stage which would increase the probability of high-amplitude variability \citep{Contreras2014}. Indeed, all these 54 YSOs show $\Delta K_{\mathrm{s}} > 0.5\,$mag. Removing them from the analysis, the null-hypothesis probability becomes $9.1\times 10^{-4}$. Hence, there remains a statistically significant difference between the two amplitude distributions, which could be due to differences in evolutionary stage or mass range between the samples, or possibly a metallicity effect related to different dust properties. However, the significant difference in the number of epochs between the \citet{Teixeira2018} and our own studies could contribute to this discrepancy, since it is expected that the \textit{measured} amplitudes may increase with the number of epochs.

Since massive Magellanic YSOs exhibit larger mass accretion rates compared with Galactic YSOs \citep{Ward2016, Ward2017}, one could have naively expected larger amplitudes for YSO variables in our sample. However, the amplitudes are either found to be similar (compared with \citealt{Borissova2016}) or smaller (compared with \citealt{Teixeira2018}). This could be due to differences in the samples, for example \textit{if} our YSO sample is slightly more evolved, it would tend to have lower accretion rates and consequently lower amplitudes. The higher gas-to-dust ratio in the LMC might also conspire to keep the amplitudes small. For the LMC this ratio is $\sim 3$ times larger than in the Galaxy \citep{Welty2012}, hence column density changes caused by inhomogeneities along the line-of-sight would lead to smaller changes in extinction. Consequently, for YSOs where extinction plays any role in causing photometric variability, a lower metallicity might actually have the effect of reducing the amplitudes. Note however, that our sample apparently contains only few objects where extinction seems to be the main driver of the variability.

\section{Summary and conclusions}
\label{sec:YSO_var_conc}
We have used NIR data from the VISTA Survey of the Magellanic Clouds (VMC) combined with observations from our open time programme to investigate photometric variability in VMC tile LMC 7\_5 in the LMC. These observations provide 15 $J$-epochs and 24 $K_{\mathrm{s}}$-epochs, covering a time-span of $\sim 6\,$years. Photometric variability was identified by applying a reduced $\chi^2$-analysis to the pawprint photometry catalogues for both bands separately. The thresholds above which a star is considered variable are defined based on distributions of the $\chi^2$-values, which depend on filter, pawprint and magnitude-range. Overall, 3062 stars of a total of 362\,425 ($\sim 0.8\,$\%) are identified as candidate variable stars. 

We selected 173 highly reliable YSO candidates identified in various \textit{Spitzer} studies. We found that 39 show variability and their light curves were studied, in terms of shape, amplitude and colour behaviour. The main results of our variability analysis are as follows: 

\begin{itemize}
  \item Of the 173 YSOs, 39 showed significant variability using the $\chi^2$-test ($\sim 23$\,\%). This fraction increases to $\sim 37$\,\% for bright YSOs ($K_{\mathrm{s}}<16\,$mag), the regime where we achieve the best sensitivity. Notably, the variability fraction increases substantially for very red sources. For $Y-K_{\mathrm{s}}>3$\,mag, $\sim 53\,$\% of the YSOs (16 of 30) show variability, but we cannot exclude selection effects.

  \item Variable YSOs exhibit, on average, redder colours than the overall YSO sample. This suggests that the variables tend to be at an earlier evolutionary stage, with possibly more significant discs or envelopes. Consequently, variability appears to be more common for the least evolved YSOs.

  \item All but three variable YSOs have moderate amplitudes in the range $0.1<\Delta K_{\mathrm{s}} < 0.6\,$mag. Compared with the amplitudes of all identified variables, YSOs show above average amplitudes. Overall, 30 variable YSOs are above the median amplitudes of the entire variable population. This implies that YSOs are dominant in samples of high amplitude variables in agreement with Galactic studies \citep{Contreras2017}.

  \item The YSO variables were classified based on the appearence of their $K_{\mathrm{s}}$-band light curves. Our sample includes: 12 eruptives, 10 faders, seven dippers, eight short-term variables (STVs), three long-period variable YSOs (LPV-YSOs) and one unclassified object. Two YSOs are members of two classes (eruptive/dipper and dipper/LPV-YSO) as they show signatures of both classes. Eruptives tend to have large amplitudes compared with the other classes. The five YSOs with the largest amplitudes are all eruptives.

  \item Long-term variability and colour changes were investigated by comparing the mean magnitudes in both bands for the early (February and March 2012), and for the most recent epochs (January and February 2018). In general, the $J$ and $K_{\mathrm{s}}$ variability properties correlate. Most YSO variables have either negligible colour changes or become bluer when fading. The results seem consistent with variability caused by unsteady accretion and/or variable grey extinction.

  \item We found two YSOs with periods in the range $2 < P < 20\,$d (where the period sensitivity of our method is best). This suggests that periodicity in this range is rare for our sample, consistent with what is expected for an early stage YSO population. The light curves of three additional YSOs indicate a possible periodicity with longer periods of $\sim 450\,$d, $\sim 490\,$d and $\sim 620\,$d.

  \item In total 40 stars from the entire sample of variables detected in the $K_{\mathrm{s}}$-band exhibit $\Delta K_{\mathrm{s}}>1\,$mag. Since most sources do not have previous {\it Spitzer}-based classifications, they were not included in the high reliability YSO sample. While only two of them belong to the variable YSO sample, 25 are spatially associated with star forming complexes. Therefore, many might in fact be massive YSOs.

  \item NIR variability is a common feature amongst YSOs in the LMC, occurring with similar frequency as in Galactic samples. The fractions of aperiodic classes (faders, dippers, eruptives) are broadly consistent with results from Galactic studies. We report fewer STVs and LPV-YSOs, likely a result to the data set properties (cadence and number of epochs).
  
  \item The observed amplitudes tend to be  similar or smaller when compared to Galactic studies. Higher amplitudes could
  naively have been expected given the larger typical accretion rates for Magellanic YSOs. On the other hand, the higher gas-to-dust ratio in the LMC would tend to decrease extinction variations and amplitudes. Intrinsic differences between the YSO samples can also not be excluded. While environmental effects could certainly play a role, we do not see the link between observed amplitudes and metallicity hypothesised above.

\end{itemize}

\section*{Acknowledgements}

This work is based on observations collected at the European Southern Observatory under ESO programmes 179.B-2003 and 0100.C-0248(A). We thank the Cambridge Astronomy Survey Unit (CASU) and the Wide Field Astronomy Unit (WFAU) for providing calibrated data products. V.Z. acknowledges studentships from ESO and the Faculty of Natural Sciences, Keele University, UK. M-R.C. and F.N. acknowledge funding support from the European Research Council (ERC) under the European Union's Horizon 2020 research and innovation programme (grant agreement No 682115). We thank the anonymous referee for their constructive comments.




\bibliographystyle{mnras}
\input{YSO_variability.bbl}




\appendix

\section{Additional Tables and Figures}
\label{app:A}

\begin{table}
	\centering
	\caption{Observational data for the $K_{\mathrm{s}}$ epochs. Included are both the VMC observations and observations from our open time programme 0100.C-0248(A).}
	\label{tab:Kepochs_data}
	\begin{tabular}{llccc}
		\hline
		UT date & Epoch & MJD & T$_{\rm exp}$ (s) & Seeing ($\arcsec$) \\
		\hline
	    2012-02-22 & Epoch\,1-K1 &  55979.068 & 375 & 0.75  \\
        2012-02-22 & Epoch\,1-K2 &  55979.074 & 375 & 0.91  \\
        2012-02-22 & Epoch\,1-K3 &  55979.081 & 375 & 0.82  \\
        2012-02-22 & Epoch\,1-K4 &  55979.088 & 375 & 0.83  \\
        2012-02-22 & Epoch\,1-K5 &  55979.095 & 375 & 0.75  \\
        2012-02-22 & Epoch\,1-K6 &  55979.101 & 375 & 0.82  \\
        2012-02-23 & Epoch\,2-K1 &  55980.080 & 175 & 0.78  \\
        2012-02-23 & Epoch\,2-K2 &  55980.087 & 175 & 0.80  \\
        2012-02-23 & Epoch\,2-K3 &  55980.091 & 175 & 0.78  \\
        2012-02-23 & Epoch\,2-K4 &  55980.094 & 175 & 0.72  \\
        2012-02-23 & Epoch\,2-K5 &  55980.098 & 175 & 0.86  \\
        2012-02-23 & Epoch\,2-K6 &  55980.101 & 175 & 0.80  \\
        2012-02-24 & Epoch\,3-K1 &  55981.054 & 375 & 0.76  \\
        2012-02-24 & Epoch\,3-K2 &  55981.061 & 375 & 0.84  \\
        2012-02-24 & Epoch\,3-K3 &  55981.067 & 375 & 0.84  \\
        2012-02-24 & Epoch\,3-K4 &  55981.074 & 375 & 0.97  \\
        2012-02-24 & Epoch\,3-K5 &  55981.081 & 375 & 0.80  \\
        2012-02-24 & Epoch\,3-K6 &  55981.087 & 375 & 0.74  \\
        2012-02-26 & Epoch\,4-K1 &  55983.115 & 200 & 0.79  \\
        2012-02-26 & Epoch\,4-K2 &  55983.119 & 200 & 0.82  \\
        2012-02-26 & Epoch\,4-K3 &  55983.123 & 200 & 0.93  \\
        2012-02-26 & Epoch\,4-K4 &  55983.127 & 200 & 0.81  \\
        2012-02-26 & Epoch\,4-K5 &  55983.131 & 200 & 0.90  \\
        2012-02-26 & Epoch\,4-K6 &  55983.135 & 200 & 0.85  \\
        2012-02-29 & Epoch\,5-K1 &  55986.054 & 375 & 0.76  \\
        2012-02-29 & Epoch\,5-K2 &  55986.061 & 375 & 0.80  \\
        2012-02-29 & Epoch\,5-K3 &  55986.067 & 375 & 0.78  \\
        2012-02-29 & Epoch\,5-K4 &  55986.074 & 375 & 0.82  \\
        2012-02-29 & Epoch\,5-K5 &  55986.080 & 375 & 0.80  \\
        2012-02-29 & Epoch\,5-K6 &  55986.087 & 375 & 0.82  \\
        2012-03-06 & Epoch\,6-K1 &  55992.066 & 375 & 1.10  \\
        2012-03-06 & Epoch\,6-K2 &  55992.073 & 375 & 1.06  \\
        2012-03-06 & Epoch\,6-K3 &  55992.080 & 375 & 0.97  \\
        2012-03-06 & Epoch\,6-K4 &  55992.086 & 375 & 0.90  \\
        2012-03-06 & Epoch\,6-K5 &  55992.093 & 375 & 0.92  \\
        2012-03-06 & Epoch\,6-K6 &  55992.100 & 375 & 0.85  \\
        2012-03-23 & Epoch\,7-K1 &  56009.010 & 375 & 0.96  \\
        2012-03-23 & Epoch\,7-K2 &  56009.017 & 375 & 0.90  \\
        2012-03-23 & Epoch\,7-K3 &  56009.024 & 375 & 0.96  \\
        2012-03-23 & Epoch\,7-K4 &  56009.030 & 375 & 0.88  \\
        2012-03-23 & Epoch\,7-K5 &  56009.037 & 375 & 0.85  \\
        2012-03-23 & Epoch\,7-K6 &  56009.044 & 375 & 0.82  \\
        2012-09-08 & Epoch\,8-K1 &  56178.345 & 375 & 0.92  \\
        2012-09-08 & Epoch\,8-K2 &  56178.352 & 375 & 0.80  \\
        2012-09-08 & Epoch\,8-K3 &  56178.358 & 375 & 0.76  \\
        2012-09-08 & Epoch\,8-K4 &  56178.365 & 375 & 0.79  \\
        2012-09-08 & Epoch\,8-K5 &  56178.372 & 375 & 0.79  \\
        2012-09-08 & Epoch\,8-K6 &  56178.378 & 375 & 0.90  \\
        2012-09-27 & Epoch\,9-K1 &  56197.284 & 375 & 0.87  \\
        2012-09-27 & Epoch\,9-K2 &  56197.291 & 375 & 0.86  \\
        2012-09-27 & Epoch\,9-K3 &  56197.298 & 375 & 0.81  \\
        2012-09-27 & Epoch\,9-K4 &  56197.305 & 375 & 0.74  \\
        2012-09-27 & Epoch\,9-K5 &  56197.312 & 375 & 0.79  \\
        2012-09-27 & Epoch\,9-K6 &  56197.318 & 375 & 0.72  \\
        2012-10-14 & Epoch\,10-K1 &  56214.326 & 375 & 0.98  \\
        2012-10-14 & Epoch\,10-K2 &  56214.332 & 375 & 0.99  \\
        2012-10-14 & Epoch\,10-K3 &  56214.339 & 375 & 0.97  \\
        2012-10-14 & Epoch\,10-K4 &  56214.346 & 375 & 0.93  \\
        2012-10-14 & Epoch\,10-K5 &  56214.352 & 375 & 0.92  \\
        2012-10-14 & Epoch\,10-K6 &  56214.359 & 375 & 0.92  \\
        2012-11-01 & Epoch\,11-K1 &  56232.241 & 375 & 0.76  \\
		\hline
	\end{tabular}
\end{table}

\begin{table}
	\centering
	\contcaption{}
	\begin{tabular}{llccc}
		\hline
		UT date & Epoch & MJD & T$_{\rm exp}$ (s) & Seeing ($\arcsec$) \\
		\hline
		2012-11-01 & Epoch\,11-K2 &  56232.248 & 375 & 0.81  \\
		2012-11-01 & Epoch\,11-K3 &  56232.254 & 375 & 0.71  \\
		2012-11-01 & Epoch\,11-K4 &  56232.261 & 375 & 0.65  \\
        2012-11-01 & Epoch\,11-K5 &  56232.268 & 375 & 0.63  \\
        2012-11-01 & Epoch\,11-K6 &  56232.274 & 375 & 0.81  \\
        2012-11-24 & Epoch\,12-K1 &  56255.158 & 375 & 0.81  \\
        2012-11-24 & Epoch\,12-K2 &  56255.165 & 375 & 0.85  \\
        2012-11-24 & Epoch\,12-K3 &  56255.173 & 375 & 0.81  \\
        2012-11-24 & Epoch\,12-K4 &  56255.180 & 375 & 0.72  \\
        2012-11-24 & Epoch\,12-K5 &  56255.187 & 375 & 0.70  \\
        2012-11-24 & Epoch\,12-K6 &  56255.193 & 375 & 0.64  \\
        2015-01-22 & Epoch\,13-K1 &  57044.029 & 375 & 1.04  \\
        2015-01-22 & Epoch\,13-K2 &  57044.036 & 375 & 1.10  \\
        2015-01-22 & Epoch\,13-K3 &  57044.042 & 375 & 1.11  \\
        2015-01-22 & Epoch\,13-K4 &  57044.049 & 375 & 1.05  \\
        2015-01-22 & Epoch\,13-K5 &  57044.056 & 375 & 0.95  \\
        2015-01-22 & Epoch\,13-K6 &  57044.062 & 375 & 0.95  \\
        2018-01-18 & Epoch\,14-K1  & 58136.069 & 480 & 0.67 \\
        2018-01-18 & Epoch\,14-K2  & 58136.077 & 480 & 0.85 \\
        2018-01-18 & Epoch\,14-K3  & 58136.085 & 480 & 0.73 \\
        2018-01-18 & Epoch\,14-K4  & 58136.094 & 480 & 0.67 \\
        2018-01-18 & Epoch\,14-K5  & 58136.102 & 480 & 0.68 \\
        2018-01-18 & Epoch\,14-K6  & 58136.110 & 480 & 0.74 \\
        2018-01-19 & Epoch\,15-K1  & 58137.119 & 480 & 0.73 \\
        2018-01-19 & Epoch\,15-K2  & 58137.128 & 480 & 0.67 \\
        2018-01-19 & Epoch\,15-K3  & 58137.137 & 480 & 0.67 \\
        2018-01-19 & Epoch\,15-K4  & 58137.145 & 480 & 0.68 \\
        2018-01-19 & Epoch\,15-K5  & 58137.153 & 480 & 0.69 \\
        2018-01-19 & Epoch\,15-K6  & 58137.162 & 480 & 0.67 \\
        2018-01-20 & Epoch\,16-K1  & 58138.080 & 480 & 0.84 \\
        2018-01-20 & Epoch\,16-K2  & 58138.089 & 480 & 1.00 \\
        2018-01-20 & Epoch\,16-K3  & 58138.097 & 480 & 1.01 \\
        2018-01-20 & Epoch\,16-K4  & 58138.106 & 480 & 0.89 \\
        2018-01-20 & Epoch\,16-K5  & 58138.114 & 480 & 0.86 \\
        2018-01-20 & Epoch\,16-K6  & 58138.123 & 480 & 0.85 \\
        2018-01-21 & Epoch\,17-K1  & 58139.078 & 480 & 0.64 \\
        2018-01-21 & Epoch\,17-K2  & 58139.087 & 480 & 0.67 \\
        2018-01-21 & Epoch\,17-K3  & 58139.095 & 480 & 0.68 \\
        2018-01-21 & Epoch\,17-K4  & 58139.104 & 480 & 0.64 \\
        2018-01-21 & Epoch\,17-K5  & 58139.112 & 480 & 0.67 \\
        2018-01-21 & Epoch\,17-K6  & 58139.121 & 480 & 0.67 \\
        2018-01-22 & Epoch\,18-K1  & 58140.082 & 480 & 0.67 \\
        2018-01-22 & Epoch\,18-K2  & 58140.090 & 480 & 0.72 \\
        2018-01-22 & Epoch\,18-K3  & 58140.098 & 480 & 0.73 \\
        2018-01-22 & Epoch\,18-K4  & 58140.107 & 480 & 0.73 \\
        2018-01-22 & Epoch\,18-K5  & 58140.116 & 480 & 0.77 \\
        2018-01-22 & Epoch\,18-K6  & 58140.124 & 480 & 0.78 \\
        2018-01-25 & Epoch\,19-K1  & 58143.046 & 480 & 0.59 \\
        2018-01-25 & Epoch\,19-K2  & 58143.055 & 480 & 0.59 \\
        2018-01-25 & Epoch\,19-K3  & 58143.063 & 480 & 0.57 \\
        2018-01-25 & Epoch\,19-K4  & 58143.071 & 480 & 0.54 \\
        2018-01-25 & Epoch\,19-K5  & 58143.080 & 480 & 0.55 \\
        2018-01-25 & Epoch\,19-K6  & 58143.088 & 480 & 0.52 \\
        2018-01-25 & Epoch\,20-K1  & 58143.152 & 480 & 0.62 \\
        2018-01-25 & Epoch\,20-K2  & 58143.160 & 480 & 0.67 \\
        2018-01-25 & Epoch\,20-K3  & 58143.169 & 480 & 0.66 \\
        2018-01-25 & Epoch\,20-K4  & 58143.177 & 480 & 0.66 \\
        2018-01-25 & Epoch\,20-K5  & 58143.186 & 480 & 0.70 \\
        2018-01-25 & Epoch\,20-K6  & 58143.194 & 480 & 0.69 \\
        2018-01-26 & Epoch\,21-K1  & 58144.071 & 480 & 0.93 \\
        2018-01-26 & Epoch\,21-K2  & 58144.080 & 480 & 0.95 \\
        2018-01-26 & Epoch\,21-K3  & 58144.088 & 480 & 1.00 \\
        2018-01-26 & Epoch\,21-K4  & 58144.096 & 480 & 0.97 \\
		\hline
	\end{tabular}
\end{table}

\begin{table}
	\centering
	\contcaption{}
	\begin{tabular}{llccc}
		\hline
		UT date & Epoch & MJD & T$_{\rm exp}$ (s) & Seeing ($\arcsec$) \\
		\hline
        2018-01-26 & Epoch\,21-K5  & 58144.105 & 480 & 0.98 \\
        2018-01-26 & Epoch\,21-K6  & 58144.113 & 480 & 0.93 \\
        2018-01-29 & Epoch\,22-K1  & 58147.171 & 480 & 0.78 \\
        2018-01-29 & Epoch\,22-K2  & 58147.179 & 480 & 0.95 \\
        2018-01-29 & Epoch\,22-K3  & 58147.188 & 480 & 0.85 \\
        2018-01-29 & Epoch\,22-K4  & 58147.196 & 480 & 0.87 \\
        2018-01-29 & Epoch\,22-K5  & 58147.205 & 480 & 0.89 \\
        2018-01-29 & Epoch\,22-K6  & 58147.213 & 480 & 0.78 \\
        2018-02-04 & Epoch\,23-K1  & 58153.053 & 480 & 0.64 \\
        2018-02-04 & Epoch\,23-K2  & 58153.061 & 480 & 0.69 \\
        2018-02-04 & Epoch\,23-K3  & 58153.070 & 480 & 0.73 \\
        2018-02-04 & Epoch\,23-K4  & 58153.105 & 480 & 0.75 \\
        2018-02-04 & Epoch\,23-K5  & 58153.114 & 480 & 0.75 \\
        2018-02-04 & Epoch\,23-K6  & 58153.122 & 480 & 0.68 \\
        2018-02-05 & Epoch\,24-K1  & 58154.146 & 480 & 0.65 \\
        2018-02-05 & Epoch\,24-K2  & 58154.154 & 480 & 0.69 \\
        2018-02-05 & Epoch\,24-K3  & 58154.163 & 480 & 0.69 \\
        2018-02-05 & Epoch\,24-K4  & 58154.171 & 480 & 0.70 \\
        2018-02-05 & Epoch\,24-K5  & 58154.180 & 480 & 0.76 \\
        2018-02-05 & Epoch\,24-K6  & 58154.188 & 480 & 0.71 \\
		\hline
	\end{tabular}
\end{table}

\clearpage
\newpage

\begin{table}
	\centering
	\caption{As Table\,\ref{tab:Kepochs_data}, but for the $J$-band.}
	\label{tab:Jepochs_data}
	\begin{tabular}{llccc}
		\hline
		UT date & Epoch & MJD & T$_{\rm exp}$ (s) & Seeing ($\arcsec$) \\
		\hline
        2011-01-28 & Epoch\,1-J1 &  55589.158  & 400 & 0.91 \\
        2011-01-28 & Epoch\,1-J2 &  55589.165  & 400 & 0.86 \\
        2011-01-28 & Epoch\,1-J3 &  55589.171  & 400 & 0.80 \\
        2011-01-28 & Epoch\,1-J4 &  55589.176  & 400 & 0.83 \\
        2011-01-28 & Epoch\,1-J5 &  55589.184  & 400 & 0.95 \\
        2011-01-28 & Epoch\,1-J6 &  55589.190  & 400 & 0.89 \\
        2011-02-23 & Epoch\,2-J1 &  55615.061  & 400 & 0.98 \\
        2011-02-23 & Epoch\,2-J2 &  55615.067  & 400 & 0.99 \\
        2011-02-23 & Epoch\,2-J3 &  55615.074  & 400 & 1.01 \\
        2011-02-23 & Epoch\,2-J4 &  55615.080  & 400 & 1.20 \\
        2011-02-23 & Epoch\,2-J5 &  55615.086  & 400 & 1.19 \\
        2011-02-23 & Epoch\,2-J6 &  55615.092  & 400 & 1.11 \\
        2012-02-23 & Epoch\,3-J1 &  55980.059  & 200 & 0.79 \\
        2012-02-23 & Epoch\,3-J2 &  55980.062  & 200 & 0.80 \\
        2012-02-23 & Epoch\,3-J3 &  55980.066  & 200 & 0.85 \\
        2012-02-23 & Epoch\,3-J4 &  55980.069  & 200 & 0.74 \\
        2012-02-23 & Epoch\,3-J5 &  55980.072  & 200 & 0.88 \\
        2012-02-23 & Epoch\,3-J6 &  55980.076  & 200 & 0.87 \\
        2012-03-18 & Epoch\,4-J1 &  56004.029  & 200 & 0.92 \\
        2012-03-18 & Epoch\,4-J2 &  56004.032  & 200 & 1.01 \\
        2012-03-18 & Epoch\,4-J3 &  56004.036  & 200 & 1.04 \\
        2012-03-18 & Epoch\,4-J4 &  56004.039  & 200 & 0.95 \\
        2012-03-18 & Epoch\,4-J5 &  56004.043  & 200 & 0.93 \\
        2012-03-18 & Epoch\,4-J6 &  56004.046  & 200 & 0.88 \\
        2018-01-18 & Epoch\,5-J1  & 58136.120   & 90  & 0.72 \\
        2018-01-18 & Epoch\,5-J2  & 58136.121   & 90  & 0.80 \\
        2018-01-18 & Epoch\,5-J3  & 58136.123   & 90  & 0.80 \\
        2018-01-18 & Epoch\,5-J4  & 58136.125   & 90  & 0.84 \\
        2018-01-18 & Epoch\,5-J5  & 58136.127   & 90  & 0.88 \\
        2018-01-18 & Epoch\,5-J6  & 58136.128   & 90  & 0.81 \\
        2018-01-19 & Epoch\,6-J1  & 58137.171   & 90  & 0.65 \\
        2018-01-19 & Epoch\,6-J2  & 58137.173   & 90  & 0.66 \\
        2018-01-19 & Epoch\,6-J3  & 58137.174   & 90  & 0.66 \\
        2018-01-19 & Epoch\,6-J4  & 58137.176   & 90  & 0.66 \\
        2018-01-19 & Epoch\,6-J5  & 58137.178   & 90  & 0.65 \\
        2018-01-19 & Epoch\,6-J6  & 58137.180   & 90  & 0.67 \\
        2018-01-20 & Epoch\,7-J1  & 58138.131   & 90  & 0.89 \\
        2018-01-20 & Epoch\,7-J2  & 58138.133   & 90  & 0.86 \\
        2018-01-20 & Epoch\,7-J3  & 58138.135   & 90  & 0.88 \\
        2018-01-20 & Epoch\,7-J4  & 58138.137   & 90  & 0.88 \\
        2018-01-20 & Epoch\,7-J5  & 58138.138   & 90  & 0.82 \\
        2018-01-20 & Epoch\,7-J6  & 58138.140   & 90  & 0.88 \\
        2018-01-21 & Epoch\,8-J1  & 58139.129   & 90  & 0.76 \\
        2018-01-21 & Epoch\,8-J2  & 58139.131   & 90  & 0.79 \\
        2018-01-21 & Epoch\,8-J3  & 58139.133   & 90  & 0.77 \\
        2018-01-21 & Epoch\,8-J4  & 58139.135   & 90  & 0.78 \\
        2018-01-21 & Epoch\,8-J5  & 58139.137   & 90  & 0.81 \\
        2018-01-21 & Epoch\,8-J6  & 58139.139   & 90  & 0.76 \\
        2018-01-22 & Epoch\,9-J1  & 58140.133   & 90  & 0.81 \\
        2018-01-22 & Epoch\,9-J2  & 58140.135   & 90  & 0.81 \\
        2018-01-22 & Epoch\,9-J3  & 58140.137   & 90  & 0.79 \\
        2018-01-22 & Epoch\,9-J4  & 58140.138   & 90  & 0.77 \\
        2018-01-22 & Epoch\,9-J5  & 58140.140   & 90  & 0.76 \\
        2018-01-22 & Epoch\,9-J6  & 58140.142   & 90  & 0.83 \\
        2018-01-25 & Epoch\,10-J1  & 58143.097   & 90  & 0.53 \\
        2018-01-25 & Epoch\,10-J2  & 58143.099   & 90  & 0.58 \\
        2018-01-25 & Epoch\,10-J3  & 58143.101   & 90  & 0.61 \\
        2018-01-25 & Epoch\,10-J4  & 58143.103   & 90  & 0.59 \\
        2018-01-25 & Epoch\,10-J5  & 58143.105   & 90  & 0.60 \\
        2018-01-25 & Epoch\,10-J6  & 58143.106   & 90  & 0.65 \\
        2018-01-25 & Epoch\,11-J1  & 58143.203   & 90  & 0.74 \\
        2018-01-25 & Epoch\,11-J2  & 58143.204   & 90  & 0.74 \\
		\hline
	\end{tabular}
\end{table}

\begin{table}
	\centering
	\contcaption{}
	\begin{tabular}{llccc}
		\hline
		UT date & Epoch & MJD & T$_{\rm exp}$ (s) & Seeing ($\arcsec$) \\
		\hline
        2018-01-25 & Epoch\,11-J3  & 58143.206   & 90  & 0.74 \\
        2018-01-25 & Epoch\,11-J4  & 58143.208   & 90  & 0.74 \\
        2018-01-25 & Epoch\,11-J5  & 58143.210   & 90  & 0.75 \\
        2018-01-25 & Epoch\,11-J6  & 58143.212   & 90  & 0.71 \\
        2018-01-26 & Epoch\,12-J1  & 58144.122   & 90  & 1.05 \\
        2018-01-26 & Epoch\,12-J2  & 58144.124   & 90  & 1.01 \\
        2018-01-26 & Epoch\,12-J3  & 58144.126   & 90  & 1.02 \\
        2018-01-26 & Epoch\,12-J4  & 58144.128   & 90  & 0.94 \\
        2018-01-26 & Epoch\,12-J5  & 58144.130   & 90  & 0.96 \\
        2018-01-26 & Epoch\,12-J6  & 58144.132   & 90  & 0.97 \\
        2018-01-29 & Epoch\,13-J1  & 58147.222   & 90  & 0.92 \\
        2018-01-29 & Epoch\,13-J2  & 58147.224   & 90  & 1.00 \\
        2018-01-29 & Epoch\,13-J3  & 58147.226   & 90  & 1.03 \\
        2018-01-29 & Epoch\,13-J4  & 58147.228   & 90  & 1.03 \\
        2018-01-29 & Epoch\,13-J5  & 58147.230   & 90  & 1.00 \\
        2018-01-29 & Epoch\,13-J6  & 58147.231   & 90  & 0.97 \\
        2018-02-04 & Epoch\,14-J1  & 58153.130   & 90  & 0.69 \\
        2018-02-04 & Epoch\,14-J2  & 58153.132   & 90  & 0.74 \\
        2018-02-04 & Epoch\,14-J3  & 58153.134   & 90  & 0.72 \\
        2018-02-04 & Epoch\,14-J4  & 58153.136   & 90  & 0.74 \\
        2018-02-04 & Epoch\,14-J5  & 58153.137   & 90  & 0.68 \\
        2018-02-04 & Epoch\,14-J6  & 58153.139   & 90  & 0.68 \\
        2018-02-05 & Epoch\,15-J1  & 58154.197   & 90  & 0.82 \\
        2018-02-05 & Epoch\,15-J2  & 58154.198   & 90  & 1.11 \\
        2018-02-05 & Epoch\,15-J3  & 58154.200   & 90  & 0.91 \\
        2018-02-05 & Epoch\,15-J4  & 58154.206   & 90  & 0.91 \\
        2018-02-05 & Epoch\,15-J5  & 58154.208   & 90  & 0.81 \\
        2018-02-05 & Epoch\,15-J6  & 58154.210   & 90  & 0.88 \\
		\hline
	\end{tabular}
\end{table}

\clearpage

\begin{table*}
	\centering
	\caption{Observational data for the 3062 identified variable star candidates. From left to right listed are the star catalogue IDs, coordinates, $YJK_{\mathrm{s}}$ magnitudes taken from the deep catalogue, band(s) in which the star was identified as variable, amplitudes in $J$ and $K_{\mathrm{s}}$, and variability types according to OGLE (if applicable). The full list is provided online.}
	\label{tab:allvar}
	\begin{tabular}{rccccccccc}
		\hline
		ID & RA\,(J2000) & Dec\,(J2000) & $Y$ & $J$ & $K_{\mathrm{s}}$ & variable band(s) & $\Delta J$ & $\Delta K_{\mathrm{s}}$ & OGLE detected \\
		& & & (mag) & (mag) & (mag) & & (mag) & (mag) & \\
		\hline
		103  & $82\fdg8384$ & $-67\fdg3631$ & 16.20 & 16.00 & 15.77 & $J$, $K_{\mathrm{s}}$ & 0.12 & 0.16 & -\\
		1430 & $82\fdg5914$ & $-67\fdg3621$ & 16.76 & 16.72 & 16.86 & $K_{\mathrm{s}}$ & 0.13 & 0.28 & -\\
		2331 & $82\fdg5529$ & $-67\fdg3613$ & 16.90 & 16.41 & 15.77 & $K_{\mathrm{s}}$ & 0.22 & 0.18 & -\\
		3548 & $82\fdg3259$ & $-67\fdg3603$ & 16.07 & 15.90 & 15.70 & $J$, $K_{\mathrm{s}}$ & 0.10 & 0.16 & -\\
		3798 & $82\fdg8175$ & $-67\fdg3601$ & 17.44 & 17.02 & 16.41 & $J$, $K_{\mathrm{s}}$ & 0.26 & 0.36 & -\\
		... & ... & ... & ... & ... & ... & ... & ... & ... & ...\\
		\hline
	\end{tabular}
\end{table*}

\clearpage

\begin{figure}
  \centering
  \includegraphics[width=\columnwidth]{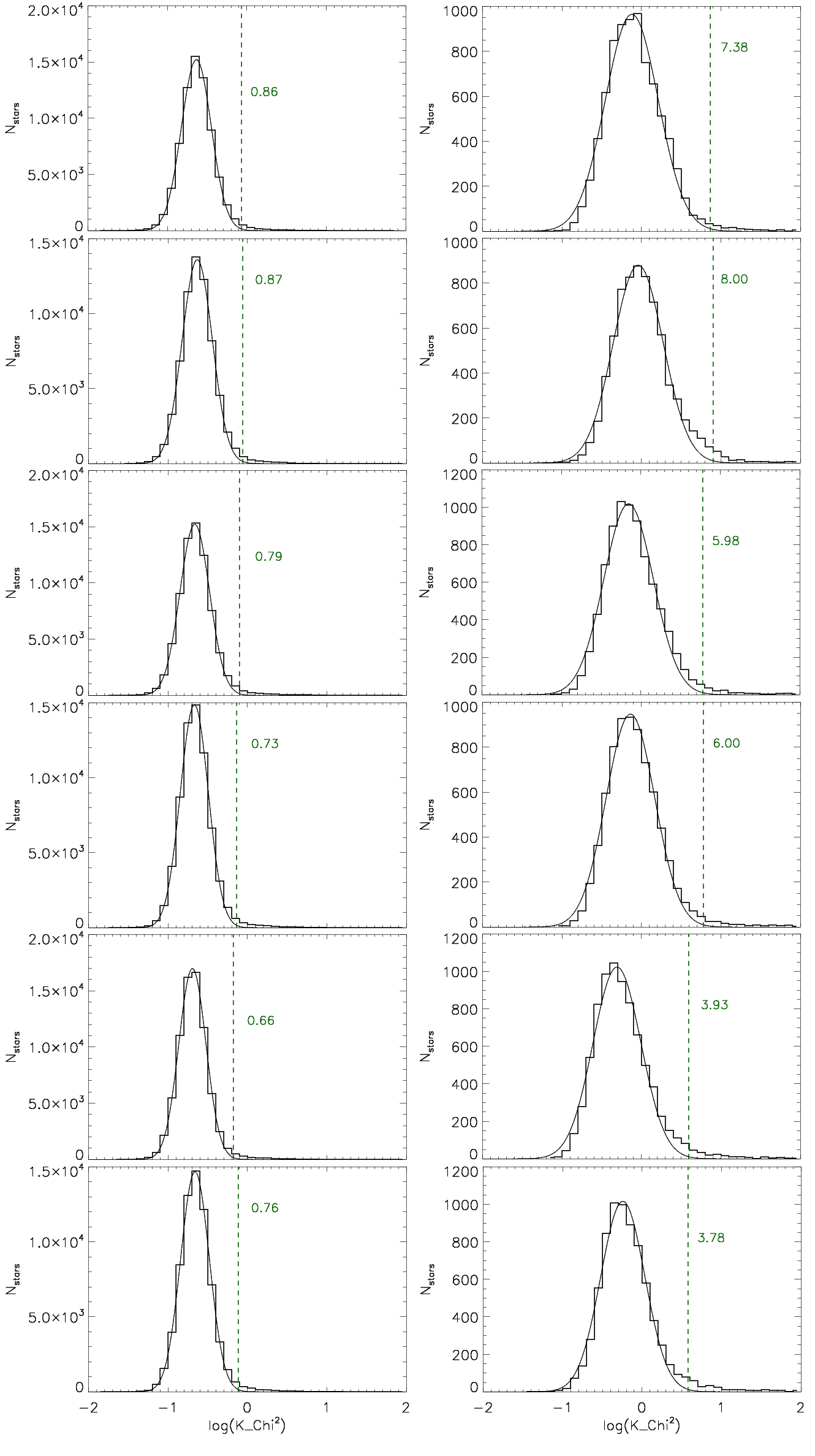}
    \caption{$\chi^2$-distributions for $K_{\mathrm{s}}$ band photometric data for Pawprint \#1 to \#6 (top to bottom row): left panels for the faint sample ($K_{\mathrm{s}} \geqslant 15.5\,$mag), right panels for the bright sample ($K_{\mathrm{s}} < 15.5\,$mag). Dashed lines and their corresponding values indicate the $3\sigma$ thresholds.}
    \label{fig:chi2_dis}
\end{figure}

\begin{figure}
  \centering
  \includegraphics[width=\columnwidth]{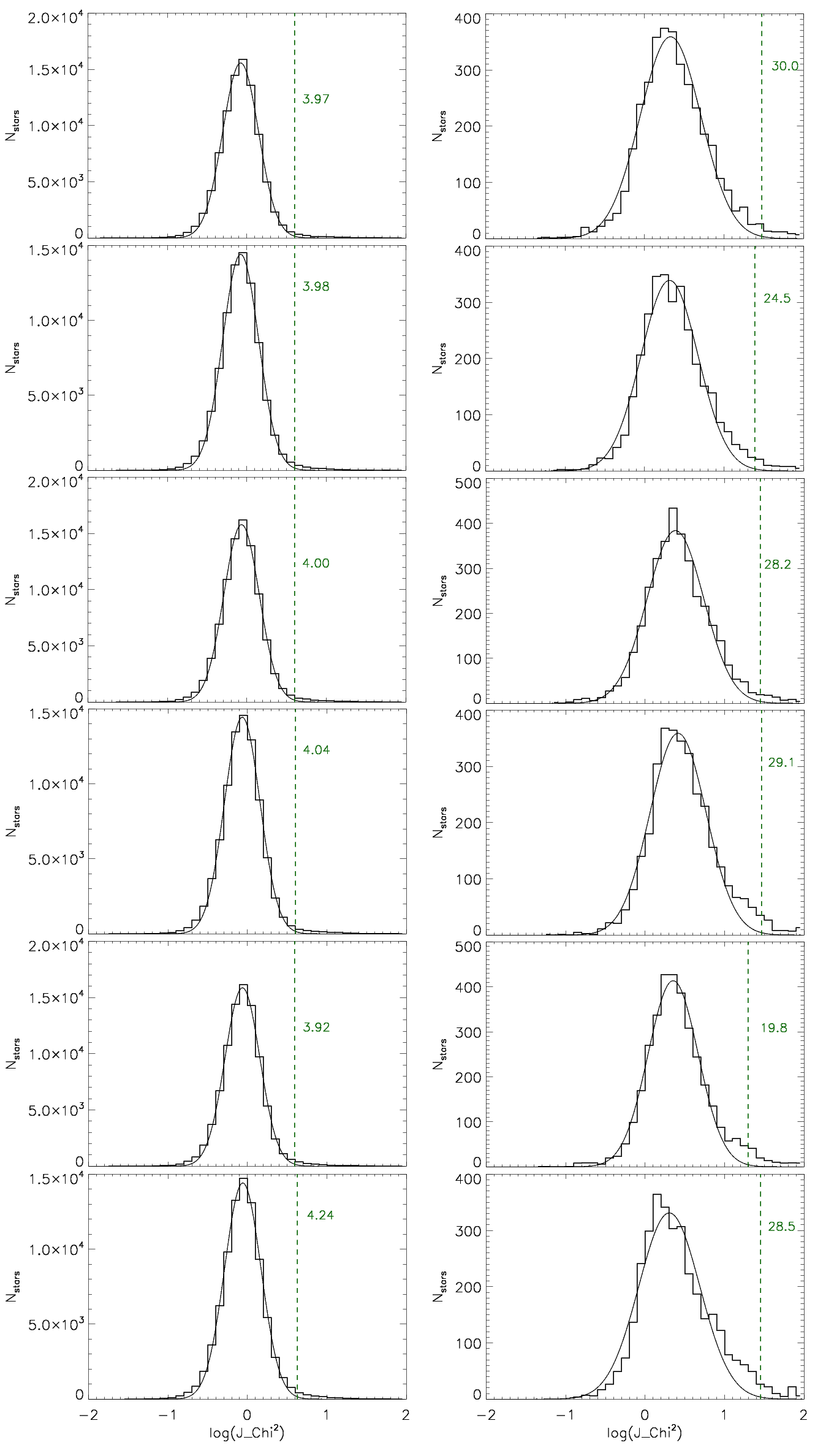}
    \caption{As Fig.\,\ref{fig:chi2_dis}, but for $J$ band photometry.
        The faint sample covers $J \geqslant 15.3\,$mag (bright sample: $J < 15.3\,$mag).}
    \label{fig:Jchi2_dis}
\end{figure}

\begin{figure*}
  \centering
  \includegraphics[width=\linewidth]{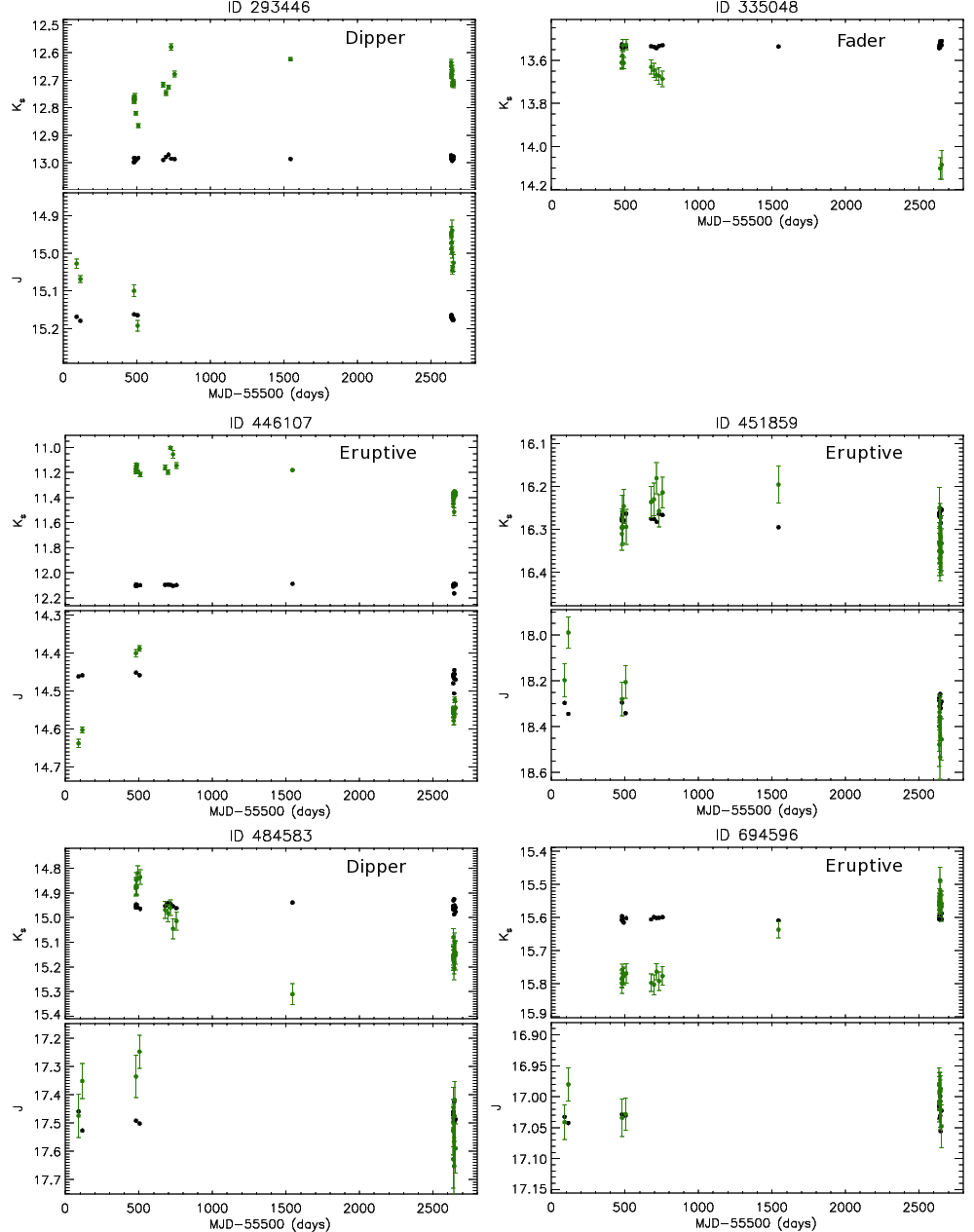}
    \caption{$K_{\mathrm{s}}$ and $J$ light curves for the identified YSO variables (green) and for a nearby comparison star (black). The ID of the YSO is indicated (as given in Table \ref{tab:ysovar}). In a few cases the $J$ light curve is missing since no $J$ band measurements are available.}
    \label{fig:lc_appendix}
\end{figure*}

\begin{figure*}
  \centering
  \includegraphics[width=\linewidth]{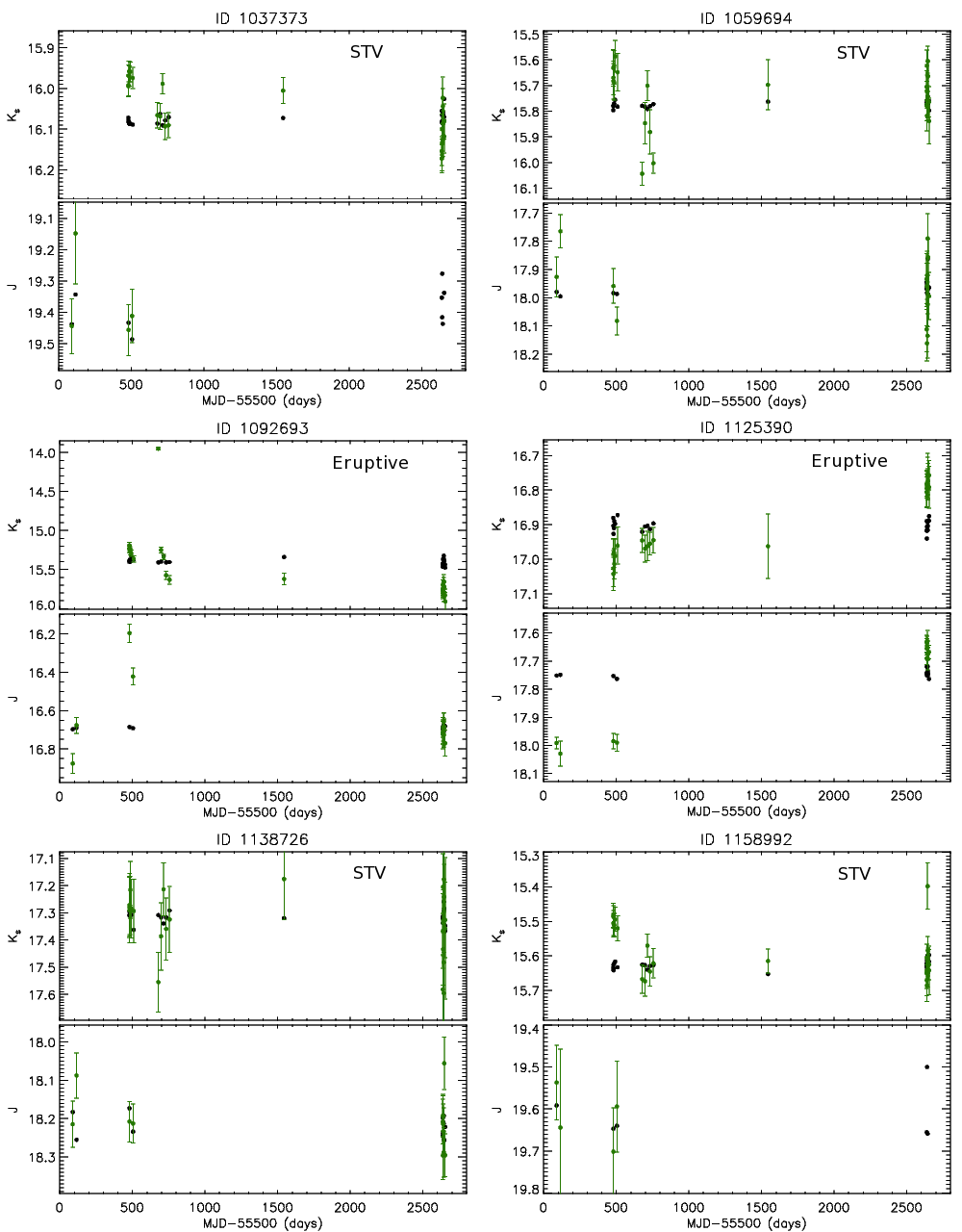}
    \contcaption{}
\end{figure*}

\begin{figure*}
  \centering
  \includegraphics[width=\linewidth]{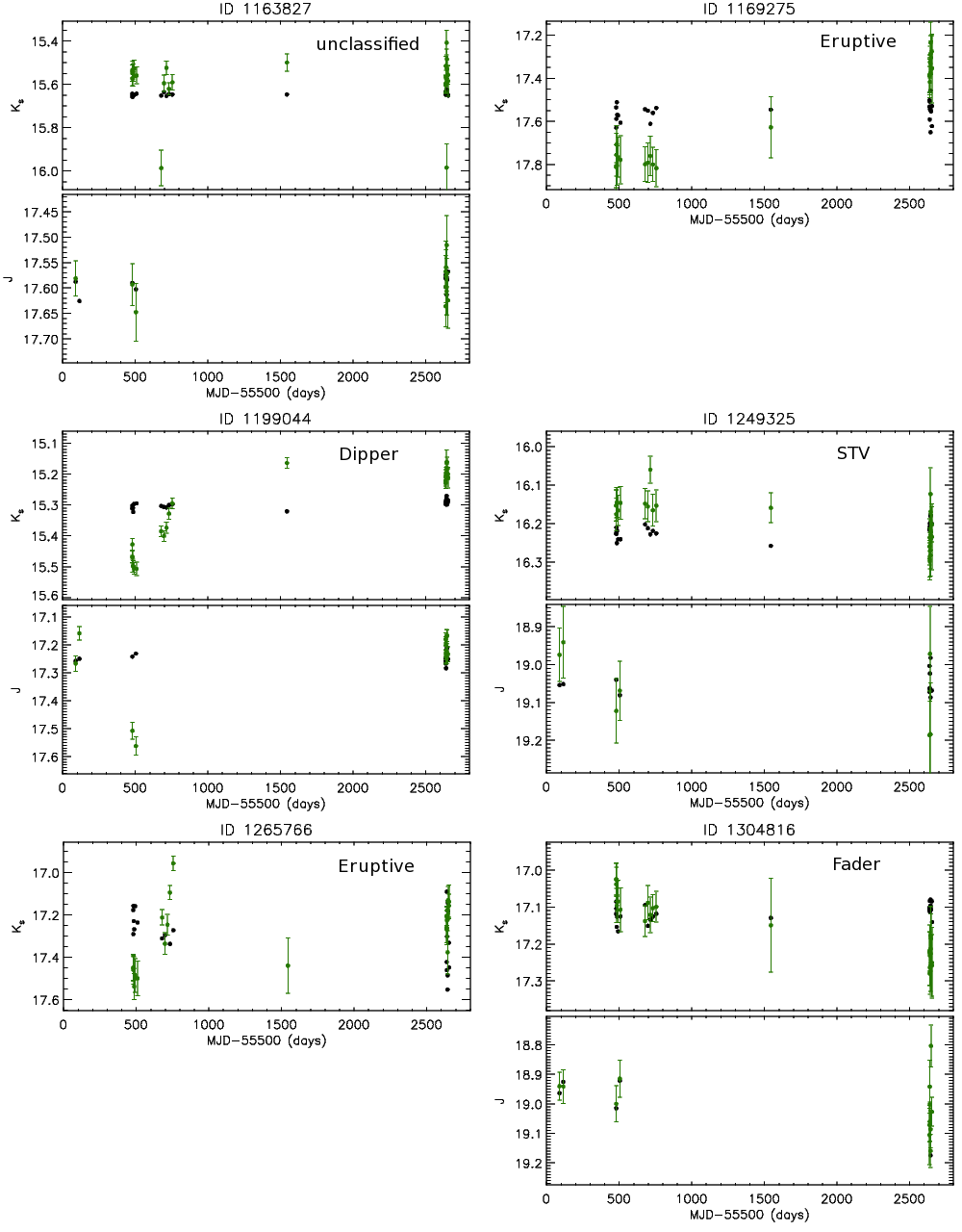}
    \contcaption{}
\end{figure*}

\begin{figure*}
  \centering
  \includegraphics[width=\linewidth]{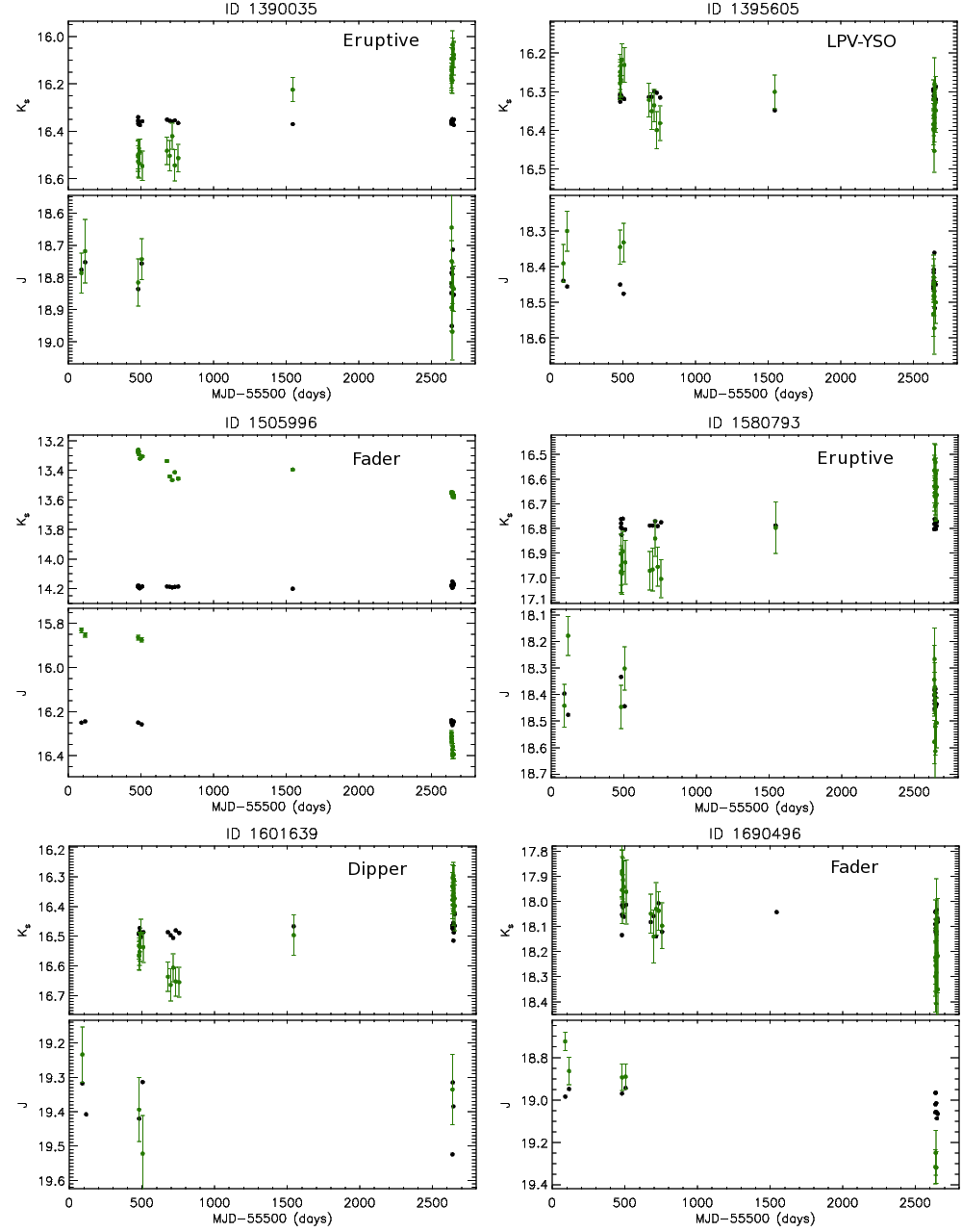}
    \contcaption{}
\end{figure*}

\begin{figure*}
  \centering
  \includegraphics[width=\linewidth]{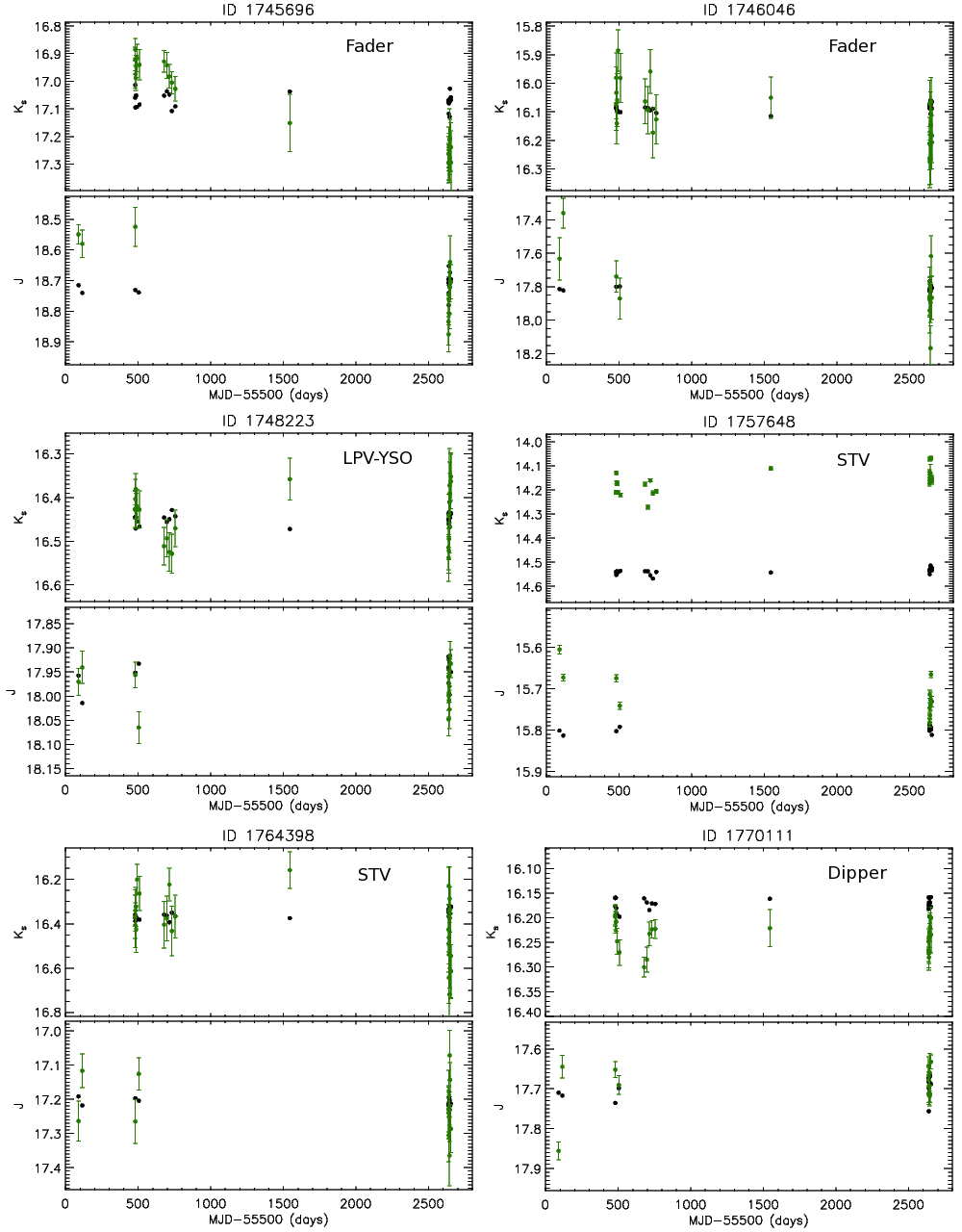}
    \contcaption{}
\end{figure*}

\begin{figure*}
  \centering
  \includegraphics[width=\linewidth]{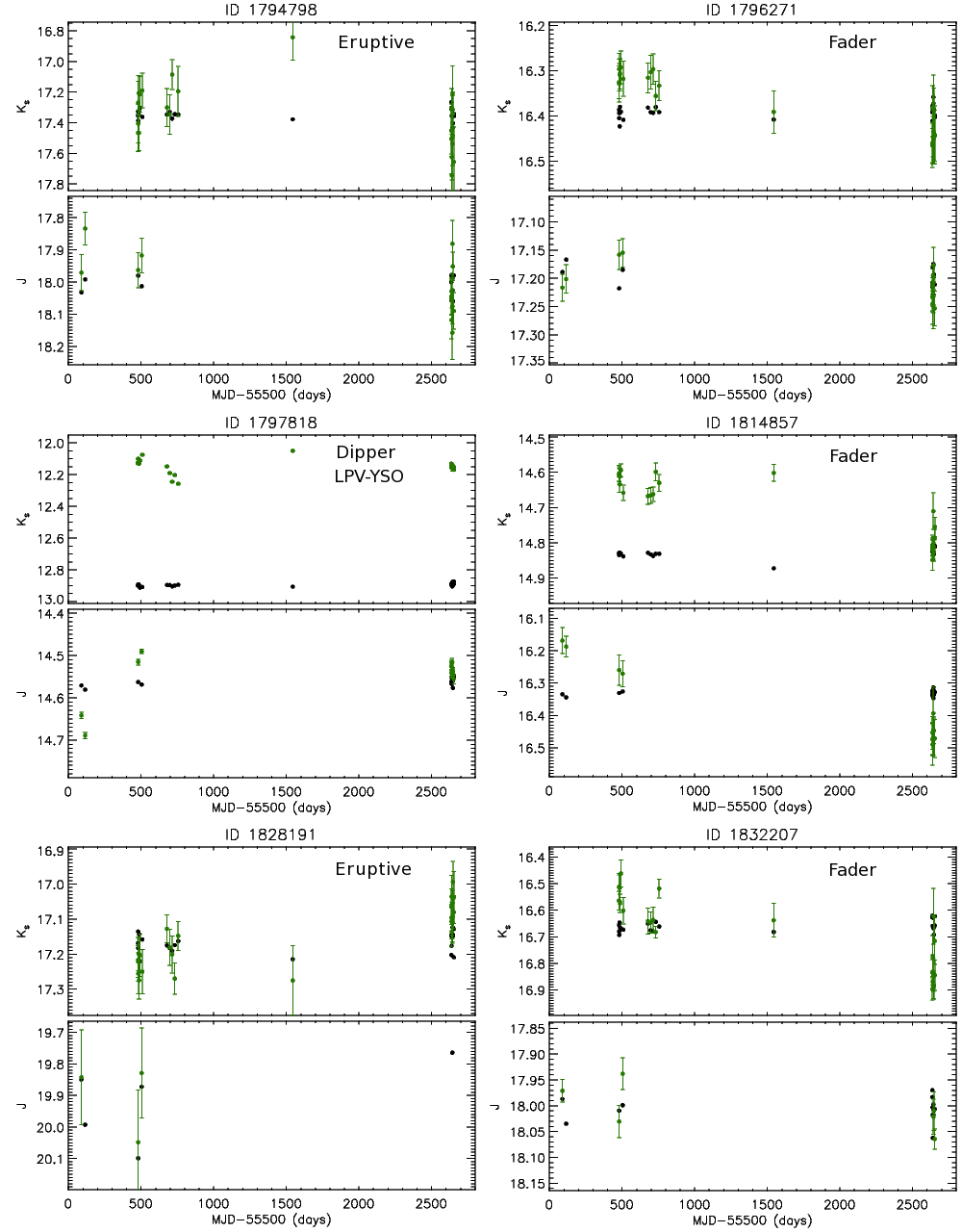}
    \contcaption{}
\end{figure*}

\begin{figure*}
  \centering
  \includegraphics[width=\linewidth]{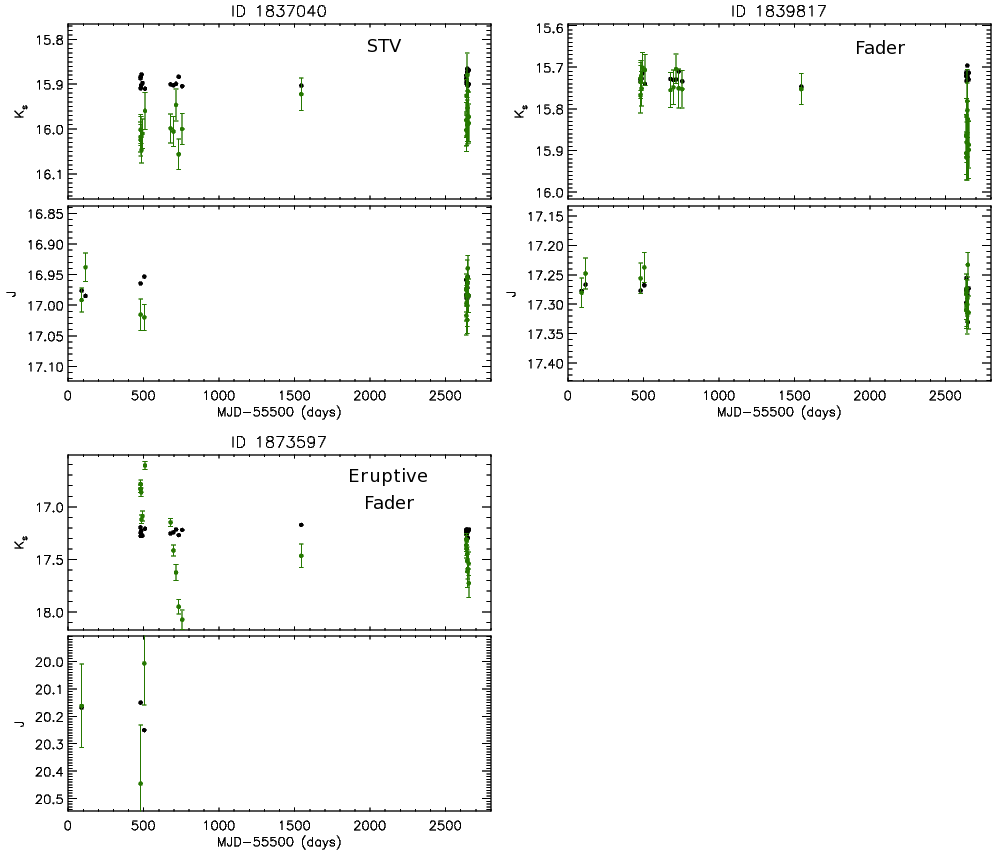}
    \contcaption{}
\end{figure*}



\bsp	
\label{lastpage}
\end{document}